\documentclass[12pt]{article}

\usepackage{float}
\usepackage{adjustbox}
\usepackage{setspace,graphicx,epstopdf,amsmath,amsfonts,amssymb,amsthm,versionPO,bm}
\usepackage{marginnote,datetime,enumitem,subfigure,rotating,fancyvrb}
\usepackage[hyperfootnotes=false]{hyperref}
\usepackage[longnamesfirst]{natbib}
\usepackage{relsize}
\usepackage{lipsum}
\usepackage[toc,page]{appendix}
\usepackage{tikz}
\usetikzlibrary{positioning}
\tikzset{main node/.style={circle,fill=blue!20,draw,minimum size=1cm,inner sep=0pt},
            }

\newcommand\blfootnote[1]{%
  \begingroup
  \renewcommand\thefootnote{}\footnote{#1}%
  \addtocounter{footnote}{-1}%
  \endgroup
}
\usdate

\excludeversion{notes}		% Include notes?
\includeversion{links}          % Turn hyperlinks on?

\iflinks{}{\hypersetup{draft=true}}

\ifnotes{%
\usepackage[margin=1in,paperwidth=10in,right=2.5in]{geometry}%
\usepackage[textwidth=1.4in,shadow,colorinlistoftodos]{todonotes}%
}{%
\usepackage[margin=1in]{geometry}%
\usepackage[disable]{todonotes}%
}
\usepackage{bbm}

\makeatletter\let\chapter\@undefined\makeatother % Undefine \chapter for todonotes

\usepackage{indentfirst} % Indent first sentence of a new section.
\usepackage{jfe}          % JFE-specific formatting of sections, etc.
\usepackage{wrapfig,lipsum,booktabs}

\usepackage{dsfont}
\usepackage{enumerate}

\newtheorem{prop}{Proposition}

\newtheorem{lem}{Lemma}
\newtheorem{defn}{Definition}

\newtheorem{clm}{Claim}

\theoremstyle{definition}
\newtheorem{examplex}{Example}

\newenvironment{example}
  {\pushQED{\qed}\examplex}
  {\popQED\endexamplex}

\begin{document}

\setlist{noitemsep}  % Reduce space between list items (itemize, enumerate, etc.)
% Use endnotes instead of footnotes - redefine \footnote command

\title{Diversity Preferences, Affirmative Action and Choice Rules}
\author{
O\u{g}uzhan \c{C}elebi\footnote{Department of Economics, Stanford University, 579 Jane Stanford Way, Stanford, CA 94305. Email: ocelebi@stanford.edu}
}

%\linespread{1.25}
% Create title page with no page number

\onehalfspacing      % Use 1.5 spacing
%\doublespacing

% Paper written using the JFE template at https://www.sharelatex.com/templates/journals

\maketitle
\vspace{-.2in}
\begin{abstract}  \noindent 
I study the relationship between diversity preferences and the choice rules implemented by institutions, with a particular focus on the affirmative action policies. I characterize the choice rules that can be rationalized by diversity preferences and demonstrate that the recently rescinded affirmative action mechanism used to allocate government positions in India cannot be rationalized. I show that if institutions evaluate diversity without considering intersectionality of identities, their choices cannot satisfy the crucial substitutes condition. I characterize choice rules that satisfy the substitutes condition and are rationalizable by preferences that are separable in diversity and match quality domains.
\end{abstract}

\blfootnote{I am grateful to Daron Acemoglu, Roberto Corrao, Federico Echenique, Glenn Ellison, Joel Flynn, Stephen Morris, Michael Ostrovsky, Parag Pathak, Alvin Roth, Tayfun S{\"o}nmez, Bumin Yenmez, Alexander Wolitzky and participants in the MIT Theory Lunch for helpful discussions and comments. First Draft: August 2022.}

% Paper written using the JFE template at https://www.sharelatex.com/templates/journals

%\clearpage

\maketitle
\clearpage
\section{Introduction}
Institutions in charge of allocating resources or hiring individuals make their decisions based on multiple criteria, such as the quality of the individuals they hire, the benefits individuals receive from the allocated resource, and the socioeconomic characteristics of these individuals. School districts in Chicago \citep{dur2018reserve} and Boston \citep{dur2020explicit} and universities in Brazil \citep{aygun2021college} prefer schools to have a diverse student body, medical authorities prefer the allocation of scarce treatments to consider equity and diversity while making sure that medical workers can receive treatment \citep{pathak2021fair,akbarpour2021economic,grigoryan2021effective} and the Indian government uses protections for historically discriminated groups when allocating government positions \citep{aygun2017large,sonmez2022affirmative}. In these settings, individuals are heterogeneous in two domains. The first is their identity; socioeconomic status for students, healthcare worker status for patients, or caste for government position applicants. The second is their score; exam scores in student assignment and government job allocation or index of clinical need in medical resource allocation. These scores may measure match quality; allocating the medical resource to a sicker individual or
a government job to a higher achieving candidate might generate more benefit or represent individual's property rights; students with higher scores might deserve the seats in selective public schools more than others. 

In many settings, institutions implement affirmative action programs where whether or not an individual is chosen depends not only on their score, but also on the composition of the identities of other chosen individuals. This shows that they care about values such as diversity and equity in addition to their preference for allocating the resource to individuals with the highest scores. This paper presents a theory of diversity preferences and studies the relationship between (i) how institutions evaluate diversity (or more generally, the composition of the chosen individuals with respect to their socioeconomic characteristics) and consider the trade-offs between diversity and scores, and (ii) how they determine the allocation rules and affirmative action policies. To this end, I develop a model in which individuals are heterogeneous in their (multidimensional) identity and their score, the institution has preferences on the scores and identities of the chosen individuals, and selects a \textit{choice rule} which determines the set of chosen individuals.

\paragraph{Rationality of Choice Rules.} I begin by establishing a connection between consumer theory and the problem of choosing sets of individuals with different identities and match qualities. I characterize the class of choice rules that can be rationalized by diversity preferences that can depend on identities of individuals in addition to their scores. Next, I study the class of choice rules that can be rationalized by preferences that are increasing in scores, or equivalently, satisfies within-group responsiveness, which is an adaptation of the responsiveness property of \cite{roth1985college} that takes into account the groups individuals belong. These characterizations are based on the \textit{congruence axiom} of \cite{richter1966revealed} and a generalization of it that incorporates an exogenous preorder, such as scores in my setting, provided by \cite{nishimura2016}. These results complement the earlier literature by focusing on the preferences behind the choice rules instead of their axiomatic properties, such as incentive compatibility and elimination of justified envy.

Using these characterizations, I devise a test of rationality that can be applied to existing or proposed allocation mechanisms. Importantly, in the market design settings studied in this paper, we know the choice rules and allocation mechanisms, and can evaluate them directly, instead of using potentially limited choice data of individual consumers. Therefore, this test can be applied to evaluate the rationality of a rule without any data and even before the rule is implemented. I apply this test to the controversial choice rule used in India between 1995 and 2020 to assign candidates to government positions. This mechanism has some important deficiencies, caused countless court cases, and was subsequently rescinded \citep{sonmez2022affirmative}. I show that this mechanism cannot be rationalized, adding to its deficiencies and illustrating the practicality of my characterization.

\paragraph{Intersectionality and Substitutes.}Next, I study the substitutes condition, an important theoretical property of choice rules that is necessary for the existence of competitive equilibria and stable matching \citep{kelso1982job,roth1984stability,hatfield2005matching}. When identities are multidimensional, institutions can evaluate diversity in multiple ways. For example, a company can look at the number of female workers and the number of black workers to determine the \textit{diversity} of their workforce, focusing on the marginal, instead of cross-sectional, distribution of identities (\textit{i.e.}, not taking into account, for example, the number of black women). Institutions and companies often highly value diversity, incorporate it explicitly into their allocation mechanisms and hiring practices, and even publish reports evaluating the diversity of their workforce. However, reports from many institutions only include the marginal distribution of their workforce.\footnote{Apple (\href{https://www.apple.com/diversity}{Apple Inclusion \& Diversity}) and Microsoft (\href{https://query.prod.cms.rt.microsoft.com/cms/api/am/binary/RE4H2f8}{Microsoft Global Diversity \& Inclusion Report 2020}) report the fraction of employees who belong to different races and genders, while MIT does the same for its student body (\href{https://ir.mit.edu/diversity-dashboard}{MIT Diversity Dashboard}). An exception is Google (\href{https://kstatic.googleusercontent.com/files/25badfc6b6d1b33f3b87372ff7545d79261520d821e6ee9a82c4ab2de42a01216be2156bc5a60ae3337ffe7176d90b8b2b3000891ac6e516a650ecebf0e3f866}{Google Diversity Annual Report 2020}), where the cross-sectional distribution of identities is reported in the intersectional hiring section. See Figure \ref{diversityfig} for examples of different diversity reporting practices.} Similarly, many affirmative action programs in legislatures have quotas for women and minorities, but these have typically evolved separately and work independently \citep{hughes2018ethnic}. These issues are the focus of the literature on \textit{intersectionality} that studies how different identities combine to create different modes of discrimination and privilege, with a particular focus on the experience of black women in the United States (\cite{crenshaw2013demarginalizing}).

Motivated by this, I study the relationship between the way multidimensional identities are evaluated by an institution and how its allocation/hiring decisions (\textit{i.e.,} choice rules) satisfy the substitutes condition. I show that if an institution value diversity without considering intersectionality, the choice rule induced by their preferences does not satisfy the substitutes condition of \cite{roth1984stability}. This shows that intersectionality is not only important from an equity perspective \citep{carvalho2022affirmative}, but also emerges as an important consideration for selection or allocation procedures to satisfy substitutes.

\paragraph{Separability in Match Quality and Diversity.}Finally, I characterize choice rules that satisfy substitutes and treat diversity and score domains separately. That is, the preferences of the institution can be represented by a utility function that is additively separable in two terms, one that depends on the scores of chosen individuals, and the other depends on the number of chosen individuals from each group. Under this representation, the preference over two individuals can depend on their scores and the representation of their groups, but not on the scores of others or the representation of other groups. Perhaps surprisingly, these rules are characterized by adaptations of three well-known properties; a choice rule is rationalizable by a utility function that is additively separable in the score and diversity domains (where the utility is increasing in scores and concave in the representation of each group) if and only if it satisfies (within-group) responsiveness \citep{roth1985college}, substitutes \citep{roth1984stability}, and the acyclicity condition of \cite{tversky1964optimal}. Then, I map existing choice rules such as quotas and reserves to this framework and show that additive separability score and diversity domains requires a unique processing order for reserve policies.

\paragraph{Related Literature.} This paper contributes to the literature on matching with affirmative action and diversity concerns initiated by \cite{abdulkadirouglu2003school} and \cite{abdulkadirouglu2005college}. \cite{kojima2012school} studies quota policies, \cite{hafalir2013effective} introduce alternative and more efficient minority reserves and \cite{ehlers2014school} generalize reserves to accommodate
policies that have floors and ceilings. \cite{dur2018reserve} and \cite{dur2020explicit} study reserves in public schools in Boston and Chicago, \cite{kamada2017stability}, \cite{kamada2018stability} and \cite{goto2017designing} study stability and efficiency in more general matching-with-constraints models. 

This paper is related to the literature on the substitutes condition  in matching markets \citep{hatfield2005matching,hatfield2010substitutes,aygun2013matching}. A closely related paper is \cite{echenique2015control}, who characterize substitutable choice rules that maximize scores conditional on achieving (or minimizing the distance with) an ideal distribution of characteristics. This paper complements theirs by considering diversity preferences that can depend on the distribution of characteristics freely (without any restrictions or reference to an ideal point) and allow flexible trade-offs between scores of individuals and the distribution of characteristics. \cite{kojima2020job} characterize all feasibility constraints that preserve substitutability and \cite{kojima2020jobtax}  complements the analysis of their previous paper by characterizing when softer pecuniary transfer policies preserve substitutability, building on the theory of discrete convex analysis \citep{murota1998discrete,murota2016discrete}. My results complement theirs by explicitly considering the multidimensional and overlapping structure of types of individuals and focusing on underlying preferences instead of constraints.

This paper builds on the literature that studies affirmative action with multidimensional and overlapping identities. \cite{kurata2017controlled} propose a mechanism that is strategy-proof and obtains student-optimal matching. \cite{aygun2021college} study the affirmative action policies where students can qualify for affirmative action in two different dimensions and show that the way overlapping identities are treated in university admissions in Brazil can cause unfairness and incentive compatibility issues. \cite{sonmez2022affirmative} demonstrate the shortcomings of the main mechanism that assigns government positions to individuals in India, where protections for overlapping domains play an important role and propose alternative mechanisms. Finally, in a recent paper, \cite{carvalho2022affirmative} study the representativeness of affirmative action policies that do not consider intersectionality. 

\cite{chan2003does}, \cite{ellison2021efficiency}, \cite{ccelebi2022priority} and \cite{celebi2021adaptive} consider a designer with an additively separable utility function in the quality and diversity domains. The analysis in Section \ref{sec:tradeoff} complements those papers by analyzing when the choice rules adopted by institutions are rationalizable by a utility function that is additively separable in the quality and diversity domains.

\section{Model}\label{section:preliminaries}
There are $N$ dimensions that represent the identities of individuals. For each $l \in \{1,\ldots,N\}$, $\Theta_{l}$ denotes the finite set of possible groups to which an individual belong in dimension $l$, $|\Theta_{l}| \geq 2$ and  $\Theta = \Theta_1 \times \ldots \times \Theta_N$.\footnote{For example, $\Theta_1$ can denote gender where $\Theta_1=\{\text{men, woman}\}$ and $\Theta_2$ can denote income where $\Theta_2=\{\text{rich, middle class, poor}\}$.} Each individual has a score $s \in \mathcal S$, where $\mathcal S \subset \mathbb R$ is a finite set of possible scores.\footnote{For simplicity, in the main text I will assume $\mathcal S \subset \mathbb R$. Appendix \ref{app:extensionscores} shows that this is without loss if the institution treats scores the same way for each identity.} $T = \Theta \times \mathcal S$ denotes all possible \textit{types} of individuals.

For individual $i$, $\theta_{l}(i)$ denotes the group of $i$ in dimension $l$, while $\theta(i) = (\theta_{1}(i),\ldots, \theta_{N}(i))$ denotes the \textit{identity} of $i$. The function $s(i)$ denotes the score of $i$ and $t(i) = (\theta(i),s(i))$ denotes the \textit{type} of $i$. $N_{\theta}(I)$ denotes the number of individuals with identity $\theta$ in $I$ and $2^{I}_x$ denotes all $x$-element subsets of $I$.

An institution chooses $q$ individuals from $I \subseteq \mathcal I$, where $\mathcal I$ denotes the set of all individuals. Formally, a choice rule is a correspondence $C: 2^{\mathcal I} \to 2^{2^{\mathcal I}}$ such that if $I \in C(I')$, then (i) $ I \in 2^{I'}_q \text{ whenever } |I'| \geq q$ (the institution fills its capacity whenever there are enough individuals) and (ii) $I = I' \text{ whenever }  |I'| < q$ (the institution chooses all individuals if there aren't enough individuals to fill the capacity).\footnote{These are referred as capacity filling choice rules.} Let $\mathcal T \equiv \cup_{q' \in \{1,\ldots, q\}} \prod_{1}^{q'} T$ denote all possible type distributions $q$ or fewer individuals can have. For $I$ with $|I| \leq q$, $\tau(I) \in \mathcal T$ denotes the types of individuals in $I$.\footnote{Formally, $\tau(I)=(t(i_1),\ldots,t(i_{q'}))$, where $q' \leq q$.} Naturally, the choices of the institution only depends on the scores and identities of individuals, in other words, if $\hat I \in C(I)$, $\tau(I') = \tau(I)$, $\tau(\tilde I) = \tau(\hat I)$ and $\tilde I \subseteq I'$, then $\tilde I \in C(I')$.

\section{Analysis}
\subsection{Rationality of Choice Rules}\label{sec:rationality}
The preferences of the institution are represented by complete preorder $\succeq$ (with the asymmetric part $\succ$ and symmetric part $\sim$) on $\mathcal T$.\footnote{As $\mathcal T$ is finite, there is an equivalent utility representation. Appendix \ref{app:utility} extends the results to that setting.} I will slightly abuse the notation and write $I \succeq I'$ instead of $\tau(I) \succeq \tau(I')$. A preference relation $\succeq$ rationalizes $C$ if
$C$ always chooses the $\succeq$-maximal sets of individuals, that is, $C(I) = \{I': I' \succeq I'' \text{ for all } I''\subseteq I\}$. Similarly, a choice rule $C$ is induced by $\succeq$ if it returns the set of $\succeq$-maximal subsets of $I$. 

This representation is very flexible and incorporates various forms of diversity preferences. It can consider the scores and identities of individuals, as well as the number of individuals with each identity. Most importantly, $\succeq$ does not need to satisfy responsiveness \citep{roth1985college}.\footnote{Responsive preferences require that for all $I$ with $I \cap \{i,i'\} = \emptyset$, $i \succ i' \implies i \cup I \succ i' \cup I$. $\succ$ restricted to singleton sets is the primitive preference (for example, priority order of a school) and preferences over sets of individuals are constructed using this order.} For example, if $I$ has more  $\theta(i)$ individuals and $I'$ has more $\theta(i')$ individuals, the institution can have both $i \cup I' \succ i' \cup I'$ and $i' \cup I \succ i \cup I$. Therefore, the identities of other chosen individuals can affect how the institution compares $i$ and $i'$, allowing the institutions' preferences to depend on the distribution of identities.

I now define a \textit{choice cycle}, which is the main axiom that characterizes the rationality of a choice rule.

\begin{defn}
$I_1,\ldots,I_n$ is a choice cycle if 
\begin{itemize}
    \item for each $i<n$, there exists an $\hat I_i$ such that $I_i \in C(\hat I_i)$ and $I_{i+1} \subset \hat I_i$,
    \item there exists  $\hat I_n$ such that $I_n \in C(\hat I_n)$, $I_{1} \subset \hat I_n$ and  $I_{1} \not \in C(\hat I_n)$. 
\end{itemize}
\end{defn}

Existence of a choice cycle corresponds to violation of the \textit{congruence} axiom of \cite{richter1966revealed}. If there is a choice cycle under $C$, then the institution has chosen $I_1$ when $I_2$ was available, $\ldots$, $I_{n-1}$ when $I_n$ is available. Therefore, $I_1$ is indirectly (weakly) revealed preferred to $I_n$. The fact that $I_n$ is chosen when $I_1$ is available and isn't chosen  means that $I_n$ is directly (strictly) revealed preferred to $I_1$, which, as Proposition \ref{prop:rationality} shows, contradicts the rationality of the choice rule. 

\begin{prop}\label{prop:rationality}
There exists $\succeq$ that rationalizes $C$ if and only if $C$ doesn't admit a choice cycle.
\end{prop}

Rationalizability does not put any restrictions on how $\succeq$ treat scores. However, in many settings, institutions prefer higher scoring individuals to lower scoring ones. I say that $I  >_{\mathcal S} I'$ if there exists a bijection $h: I \to I'$ such that for all $i \in I$, $\theta(i) = \theta(h(i))$, $s(i) \geq s(h(i))$ and at least one of the inequalities is strict. $\succeq$ is increasing in scores if $I  >_{\mathcal S} I'$ implies $I \succ I'$. Although preferences that incorporate diversity considerations can take into account the representation of each group and don't need to be responsive, it is reasonable that they are responsive to scores conditional on the groups of individuals.\footnote{\cite{echenique2015control} defines a similar notion they call \textit{within-type compatibility} for choice functions that requires that a lower scoring individual should not be chosen over a higher scoring individuals if they belong to the same group.}
\begin{defn}
   $\succeq$ is within-group responsive if for all $i$, $i'$ and $I$ with $\{i,i'\}  \cap I = \emptyset$, $\theta(i) = \theta(i')$ and $s(i) > s(i')$, we have $i \cup I \succ i' \cup I$.
\end{defn}
Within-group responsiveness interprets the scores as the primitive preference between individuals with the same identity, and requires that the preferences over sets of individuals are responsive to this preference.\footnote{An alternative approach would be to take $\succ$ restricted to singleton sets as the primitive preference (denoted by $\succ_R$) and define within-group responsiveness using $\succ_R$ instead of scores. Appendix \ref{app:extensionscores} adopts that approach and shows that when $\succ_R$ treats the scores the same way for each group, the properties are equivalent.} The following proposition formalizes the connection between within-group responsives and increasing in scores properties.

\begin{prop}\label{prop:increasingresponsiveness}
$\succeq$ is increasing in scores if and only if it is within-group responsive.
\end{prop}
The following generalization of choice cycles characterizes the choices that can be rationalized by within-group responsive preferences.

\begin{defn}
$I_1,\ldots,I_n$ is a score-choice cycle if 
\begin{itemize}
    \item for each $i<n$, either (i) there exists an $\hat I_i$ such that $I_i \in C(\hat I_i)$ and $I_{i+1} \subset \hat I_i$, or (ii) $I_i >_{\mathcal S} I_{i+1}$.
    \item either (i) there exists  $\hat I_n$ such that $I_n \in C(\hat I_n)$, $I_{1} \subset \hat I_n$ and $I_{1} \not \in C(\hat I_n)$, or (ii) $I_{n} >_{\mathcal S} I_{1}$.
\end{itemize}
\end{defn}

A score-choice cycle has the additional requirement that higher-scoring individuals must be chosen before lower-scoring ones. It is adapted from \cite{nishimura2016} who generalize the characterization of \cite{richter1966revealed} to settings with an exogenous order. Applying their result, we can characterize the choice rules that are rationalizable by a utility function (and a preference relation) that is increasing in scores.

\begin{prop}\label{prop:monotonicity}
$C$ doesn't admit a score-choice cycle if and only if there exists a preference relation $\succeq$ that rationalizes $C$ and is increasing in $\mathcal S$.
\end{prop}

Propositions \ref{prop:rationality} and \ref{prop:monotonicity} leverage results from consumer theory to characterize the choice rules which are rationalizable by diversity preferences, that is, preferences that consider the composition or distribution of identities of individuals. From a practical perspective, there is an important difference between the settings studied by consumer theory and market design. In many applications, we not only observe the choices made by institutions but we actually know the choice rule. We are not evaluating the rationalizability of a series of decisions made by a consumer, but the rationalizability of a single decision (the choice rule) made by the institution. Thus, Propositions \ref{prop:rationality} and \ref{prop:monotonicity} can be used directly to test the rationality of choice rules without relying on any data and even before the choice rule is actually used. 

I apply this test to the choice rule mandated by the Supreme Court of India, which has been used for 25 years and was recently rescinded due to its shortcomings. I define this choice rule ($C_{S}$) when there are two groups in two dimensions.\footnote{See \cite{sonmez2022affirmative} for the more general definition of this choice rule, as well as a detailed description and analysis of the setting.} $\Theta_1 = \{g,r\}$ where $g$ denotes the general population and $r$ denotes reserve-eligible population (individuals who belong to underrepresented castes). $\Theta_2 = \{m,w\}$ denotes the gender of the individuals. For $x \in \{g,r,m,w\}$, $I^x$ denotes the set of individuals in a given category. In this setting, $C_S$ is characterized by $4$ integers, denoting the numbers of reserved positions $r$, positions open to everyone $o$, reserve positions protected for woman, $r_w \leq r$ and open positions protected for women $o_w \leq o$, and proceeds as follows.

\begin{quote}
    \textbf{Supreme Court Mandated Choice Rule $C_S$}\\
\textbf{Step 1}: Define $\mathcal M$ as the set of reserve-eligible candidates who are among the $o$ highest scoring individuals in the population.\\
\textbf{Step 2}: Assign $o_w$ positions to $o_w$ highest scoring women in $I^g \cup \mathcal M$.\footnote{If there are fewer than  $o_w$ or $r_w$ women considered at Steps 2 or 4, then remaining positions are assigned to highest scoring men who are considered at that stage.}\\
\textbf{Step 3}: Assign remaining $o-o_w$ positions to $o-o_w$ highest scoring previously unassigned individuals in $I^g \cup \mathcal M$.\\
\textbf{Step 4}: Assign $r_w$ positions to $r_w$ highest scoring previously unassigned woman in $I^r$.\\
\textbf{Step 5}: Assign $r-r_w$  positions to highest scoring previously unassigned individuals in $I^r$.
\end{quote}

$\mathcal M$ is referred to as the \textit{meritorious reserve candidates}. $C_S$ has some important shortcomings. It doesn't satisfy \textit{no justified envy}; a reserve-eligible individual can score higher than a reserve-ineligible individual and fail to receive a position while the reserve-ineligible individual receives one, which is against the philosophy of affirmative action. Moreover, that individual would have obtained the position by not disclosing their reserve eligibility, violating \textit{incentive compatibility}. The next example shows that $C_S$ admits a score-choice cycle and therefore cannot be rationalized by a preference that is increasing in scores.

\begin{example}\label{ex:SCIirrational}
$I = \{m_1^g,w_1^r,m_1^r,w_1^g\}$, where $m$ ($w$) are men (women) and superscripts denote groups of individuals. Capacity and reserves are $o=2$, $r=1$, $o_w=1$ and $r_w=1$.  Individuals' scores are given by 
$$s(m_1^g) > s(w_1^r) > s(m_1^r) > s(w_1^g)$$

$\mathcal M = \{w_1^r\}$ and $m_1^g$ and $w_1^r$ are assigned the open positions.  The only remaining reserve-eligible individual, $m_1^r$, is chosen as there is no reserve-eligible woman candidate. Therefore, $\{m_1^g,w_1^r,m_1^r\}$ is chosen when $\{m_1^g,w_1^r,w_1^g\}$ is available.

Now replace $m_1^r$ with $\tilde m_1^r$ where $s(\tilde m_1^r) \in (s(m_1^g),s(w_1^r))$. Thus,  $\{m_1^g,w_1^r, \tilde m_1^r\} >_{\mathcal S}  \{m_1^g,w_1^r,m_1^r\}$. Then, $\mathcal M = \{\tilde m_1^r\}$. Since one of the open positions is reserved for woman, in the first stage $m_1^g$ and $w_1^g$ (who is the only woman eligible at this stage) are chosen for the open positions.  In the second stage, the only remaining reserve-eligible woman, $w_1^r$ is chosen. Therefore, $\{m_1^g,w_1^g,w_1^r\}$ is chosen when $\{m_1^g,w_1^r, \tilde m_1^r\}$ is available, creating the following score-choice cycle:
$$\{m_1^g,w_1^g,w_1^r\}, \{m_1^g,w_1^r, \tilde m_1^r\}, \{m_1^g,w_1^r,m_1^r\}$$
\end{example}

Therefore, $C_S$ cannot be rationalized by preferences that are increasing in scores. Similar to the other two issues, these cycles exist because of the way meritorious reserve candidates are processed, and give yet another reason why the rule is rescinded. In Appendix \ref{sec:sciwithoutscores}, I show that $C_S$ also admits a choice cycle and isn't rationalizable even with preferences that aren't increasing in scores. Appendix \ref{app:privilege} extends the analysis to characterize the choice rules that can be rationalized by utility function that is increasing in the representation of certain groups (such as the target groups of affirmative action), based on the monotonicity axiom introduced by \cite{aygun2021college} and illustrates that we can consider other domains other than scores while characterizing preferences.

\subsection{Intersectionality and Substitutes}\label{sec:intersectionality}
This section analyzes the relationship between how institutions evaluate diversity and substitutability of their choice rules. I start the analysis with individuals who are homogeneous in terms of quality, assuming $|\mathcal S| = 1$ and suppressing scores. Appendix \ref{subsec:hetero} allows $|S|>1$, interprets scores as inverse salaries and extends the results using the gross substitutes property of \cite{kelso1982job}.

\begin{figure}[t]
    \centering
  \includegraphics[scale=0.32]{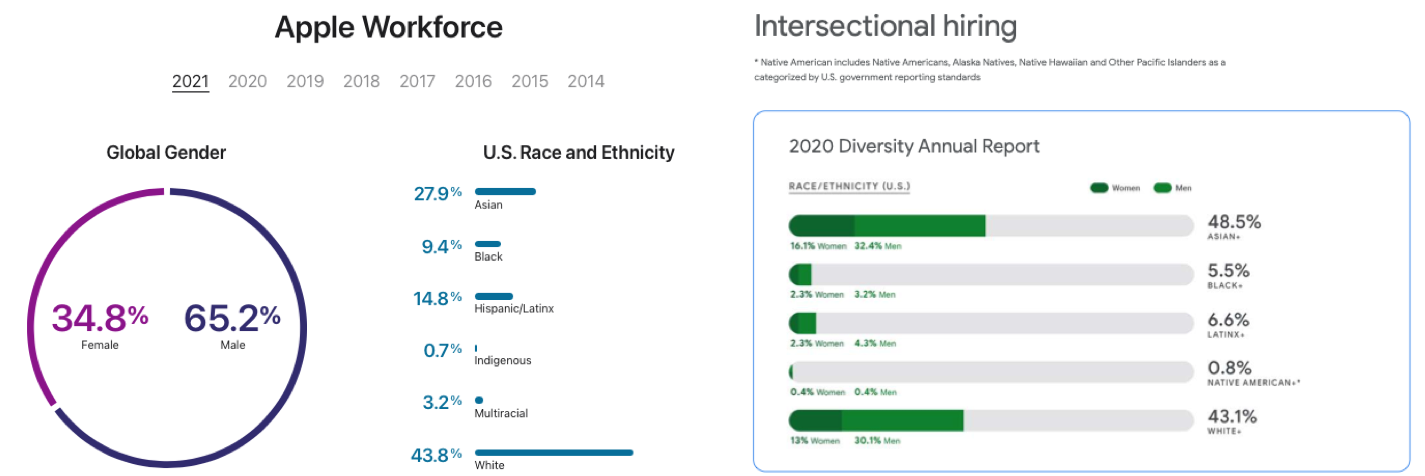}
    \caption{\small \textbf{Diversity Reporting Practices of Apple and Google.} Apple (left pane) provides a breakdown of marginal distribution of workers in race and gender without any information on the cross-sectional distribution. Google (right pane) reports the percentages of men and women for each race.}
    \label{diversityfig}
\end{figure}

Given $I$, $M_{l}(I)$ returns the number of individuals of each group in dimension $l$ and  $M(I) = (M_1(I),\ldots,M_N(I))$.\footnote{Continuing the example, if $I = \{i,j\}$ where $\theta_1(i) = \theta_1(j) = \textit{men}$, $\theta_2(i) = \textit{rich}$ and $\theta_2(j) = \textit{poor}$, then $M_1(I) = (2,0)$, $M_2(I) = (1,0,1)$ and $M(I) = \{(2,0),(1,0,1)\}$.} $M(I)$ is the \textit{marginal distribution} of $I$, as it returns the number of individuals that belong to each group in each dimension, but does not have any information about the cross-sectional distribution of groups. Figure \ref{diversityfig} presents the diversity reporting practice of Apple (which reports only marginal distributions in gender and race) and Google (which reports the gender breakdown for each race).

\begin{defn}\label{intersectionality:def}
    $\succeq$ doesn't consider intersectionality if for all $I$ and $I'$ with $M(I) = M(I') $, $I \sim I'$.
\end{defn}

Observe that if the identity is one dimensional ($|N|=1$), the marginal distribution is sufficient to determine the composition of identities in a group. Therefore, intersectionality matters when $|N|\geq2$.
    
\begin{defn}
Let $\tilde I \subseteq I' \subset I$. $C$ satisfies the substitutes condition if for all $\tilde I$ with $\tilde I \subseteq \hat I \in C(I)$, there exists $\bar I$ such that $\bar I \in C(I')$ and $\tilde I \subseteq \bar I$.
\end{defn}

This condition is the generalization of the substitutes condition of \cite{roth1984stability} to choice correspondences.\footnote{When $C$ is a choice function (\textit{i.e.}, $C(I)$ is singleton for all $I$), this condition is equivalent to the following: If $i \in C(I)$ and $I' \subseteq I$, then $i \in C(I')$. A similar generalization is employed by \cite{kojima2020jobtax} for a model with salaries.} Substitutes condition states that whenever $I^*$ is chosen from $I$, then $I^*$ is also chosen from any set $I' \subset I$ that includes it. The following example illustrates the relationship between intersectionality and the substitutes condition. Next section characterizes a large class of preferences that consider intersectionality and satisfy substitutes.

        \begin{example}\label{example:intersectionality}
    $I = \{i_1,\ldots,i_8\}$ and $q = 4$. The identities of individuals are as follows:

     \begin{center}
\begin{tabular}{ccc}
&$\Theta_1$&$\Theta_2$\\
\hline
$i_1,i_5$&$1$&$1$\\
$i_2,i_6$&$1$&$2$\\
$i_3,i_7$&$2$&$1$\\
$i_4,i_8$&$2$&$2$\\
\end{tabular}
\end{center}
\vspace{1em}
The institution has preferences $\succeq$ that doesn't consider intersectionality and strictly prefers to have exactly $2$ individuals from all $4$ groups to any other distribution. Let $C_{\succeq}$ denote the choice rule induced by $\succeq$. $\hat I = \{i_1,i_5,i_4,i_8\}$ has two individuals from each group and, therefore, $\hat I \in C_{\succeq}(I)$. However, if the set of available agents are $I \setminus \{i_1,i_5\}$, then the only set that has two individuals from each group is $\tilde I = \{i_2,i_6,i_3,i_7\}$, and $C_{\succeq}(I \setminus \{i_1,i_5\}) = \tilde I$. As $\hat I \in C(I)$ but $\hat I \not \in C(I \setminus \{i_1,i_5\})$, $C_{\succeq}$ doesn't satisfy the  substitutes condition.
    \end{example}

In Example \ref{example:intersectionality}, the institution evaluates diversity by marginal distributions which causes $\{i_1,i_5\}$ and $\{i_4,i_8\}$ to become \textit{complements}: when $\{i_4,i_8\}$ is chosen, there are two individuals from group $2$ in both dimensions and individuals who belong to group $1$ in both dimensions become more desirable. Therefore, when $\{i_1,i_5\}$ is not available, choosing $\{i_4,i_8\}$ cannot be optimal, as no two individuals from $\{i_2,i_3,i_6,i_7\}$ can complement $\{i_4,i_8\}$ to achieve the preferred distribution. 

In general, it is possible that the institution doesn't put any weight on diversity; it can be indifferent between all possible distributions. Therefore, I make a minimal assumption that makes diversity preferences non-trivial and assume that the most preferred distribution doesn't completely exclude a group from being chosen. Formally, $M(I)$ is \textit{at boundary} if $I$ has no individual from some group, \textit{i.e.}, there exists $l$ and $\hat \theta \in \Theta_l$ such that $\theta_l(i) \neq \hat \theta$ for all $i \in I$. Conversely, $M(I)$ is \textit{interior} if it isn't at boundary,  \textit{i.e.}, if $I$ has at least one individual from each group.

\begin{defn}\label{incompatibilityslide}
$\succeq$ \textbf{values diversity} if there isn't any $I$ such that $M(I)$ is at boundary and $I \succeq I'$ for all $I'$.
\end{defn}

This is a reasonable assumption for diversity preferences; it requires that the institution values diversity and prefers to choose at least one individual from each group, but puts no other restrictions on how it values different compositions of individuals. 

\begin{prop}\label{incompatibility:prop}
Suppose that $|N|\geq2$, $\succeq$ values diversity, doesn't consider intersectionality and induces $C_{\succeq}$. Then $C_{\succeq}$ doesn't satisfy the substitutes condition.
\end{prop}

Proposition \ref{incompatibility:prop} shows that the logic of Example \ref{example:intersectionality} is indeed much more general. Evaluating diversity with the marginal distributions create complementarities between certain types, and considering intersectionality is not only crucial from an equity standpoint but also necessary to satisfy the substitutes condition.

\subsection{Separable Utility Representation}\label{sec:tradeoff}
The analysis in the previous sections was silent on possible trade-offs between the quality and diversity domains. This section characterizes the choice rules that satisfy substitutes and are induced by a class of preferences that treat diversity and quality domains separately. Separability is a reasonable property to study; in various settings, the contribution of an individual to an institution is independent from their identity, and the institution prefers having a diverse body due to equity concerns, not because it would boost productivity. In what follows, I will show that three well-known properties adapted to my setting characterize the choice rules that can be rationalized by the following utility function
    \begin{equation}
        U(I) = \sum_{i \in I} u(s(i)) + \sum_{\theta \in \Theta} h_{\theta }(N_{\theta}(I))
    \end{equation}
where $ u(s(i))$ is the benefit the institution receives from allocating the resource to an individual with score $s(i)$, while $h_{\theta }(N_{\theta}(I))$ is the benefit the institution receives from choosing $N_{\theta}(I)$ individuals with each identity $\theta$. The utility from the score domain is obtained by summing the  $ u(s(i))$ over all chosen individuals, the diversity utility is obtained by summing  $h_{\theta }(N_{\theta}(I))$ over all identities and the utility of the institution is the sum of these two terms.

For this section, I assume that if $I' \in C(I)$ and $I'' \in C(I)$, then $I'$ is equivalent to $I''$, that is, $\tau(I) = \tau(I')$. This means that although $C$ is still a choice correspondence since there can be many individuals with the same type, it is actually a choice function if we restrict attention to equivalence classes $\mathcal T$. 

Given $C$, I construct the following binary relation $>_C$:
\begin{equation*}
    \begin{split}
        I \cup \{j\} \in C(I \cup \{j,k\})& \text{ and } I \cup \{k\} \not \in C(I \cup \{j,k\})\\
&\implies (s(j),\theta(j),N_{\theta(j)}(I \cup \{j\})) >_C (s(k),\theta(k),N_{\theta(k)}(I \cup \{k\}))
    \end{split}
\end{equation*}
 
$>_C$ is the revealed preference relation over pairs of individuals induced by $C$. $(s,\theta,n) >_C (s',\theta',n')$ indicates that a $\theta$ individual with score $s$ is chosen with $n-1$ other $\theta$ individuals instead of a $\theta'$ individual with score $s'$ with $n'-1$ other $\theta'$ individuals. Let $Q = \{1,\ldots,q\}$ and $D = \Theta \times Q $ denote the set of all $(\theta,n)$ with generic element $d \in D$.

\begin{defn}
Given a binary relation $>$, a collection 
\begin{equation*}
    \begin{split}
        (s_1,d_1) & > (s_1',d_1')\\
        (s_2,d_2) & >  (s_2',d_2')\\
        &\hspace{0.5em} \vdots\\
        %(s_{m-1},d_{m-1}) &\succ (s_m',d_m), (s_m,d_m)\\
        (s_m,d_m) & > (s_m',d_m')
    \end{split}
\end{equation*}
is a \textit{cycle} if $(s_1',\ldots,s_m')$ is a permutation of  $(s_1,\ldots,s_m)$ and  $(d_1',\ldots,d_m')$ is a permutation of  $(d_1,\ldots,d_m)$.
\end{defn}

This definition is due to \cite{tversky1964optimal} (see also \cite{scott1964measurement,adams1965elements}) and is used to characterize preferences that admit an additively separable utility representation. The existence of a cycle under $>_C$ means that the evaluation of the diversity and quality domains are connected, since $\{(s_i,d_i)\}_{i \leq m}$ and $\{(s_i',d_i')\}_{i \leq m}$ are formed from the same scores and diversity levels, but $\{(s_i,d_i)\}_{i \leq m}$ are revealed strictly preferred to $\{(s'_i,d'_i)\}_{i \leq m}$ for all $i$.

\begin{defn}
    $C$ satisfies \textbf{acyclicity} if there are no cycles under $>_C$.
\end{defn}

Acyclicity of $>_C$ rules out any connection between the diversity and score domains. I also adapt the definition of within-group responsiveness to choice rules, replacing $\succeq$ with $C$.

\begin{defn}
     $C$ satisfies \textbf{within-group responsiveness} if for all $i$ and $j$ with $\theta(i) = \theta(j)$ and $s(i) > s(j)$, there does not exist $\hat I \subseteq I$ such that $\hat I \cap \{i,j\} = \emptyset$, $\hat I \cup \{j\} \in C(I)$ and  $\hat I \cup \{i\} \not \in C(I)$.
\end{defn}

Within-group responsiveness makes sure the choice rule is responsive to scores in the sense that higher scoring individuals are chosen before lower scoring individuals. If within-group responsiveness fails, then $\hat I \cup \{j\}$ is chosen while $\hat I \cup \{i\}$ is not and the choice rule is not responsive to scores. The following result characterizes the class of choice rules that can be induced by a utility function that is separable in the diversity and quality domains.

\begin{prop}\label{GSrepresentation}
    If $C$ satisfies substitutes, within-group responsiveness and acyclity, then there exist increasing $u$ and concave $\{h_{\theta}\}_{\theta \in \Theta}$ such that
    \begin{equation}\label{eq:seputility}
        U(I) = \sum_{i \in I} u(s(i)) + \sum_{\theta \in \Theta} h_{\theta }(N_{\theta}(I))
    \end{equation}
    where $U$ rationalizes $C$. For each increasing $u$ and concave $\{h_{\theta}\}_{\theta \in \Theta}$, $C_U$ satisfies substitutes, within-group responsiveness and acyclity.
\end{prop}

Proposition \ref{GSrepresentation} shows that a choice rule can be induced by a utility function given in Equation \ref{eq:seputility} if and only if it satisfies substitutes \citep{roth1984stability}, within-group responsiveness \citep{roth1985college} and acyclity \citep{tversky1964optimal}. In addition to additive separability, $U$ has a preference for diversity as the functions $h_{\theta}$ are concave, which means that the marginal benefit of choosing an individual with a given identity is weakly decreasing in the number of such individuals, representing a preference against choosing many individuals with the same identity.

To prove Proposition \ref{GSrepresentation}, I show that the incomplete binary relation $>_C$ can be represented by an additively separable utility function $u(s) + h(\theta,n)$ over $(s,\theta,n)$ tuples, where $h(\theta,n)$ represents the benefit of adding the $n$'th individual with identity $\theta$ and $u(s)$ is increasing in $s$. At this point, $u$ and $\{h(\theta,n)\}_{\theta \in \Theta}$ represent $C$ for decisions between pairs of individuals, but not necessarily for decisions over sets of individuals. Then I show that, under substitutes, we can construct concave $h_{\theta}$ using $h(\theta,n)$ where $h_{\theta}(n)$ represents the benefit received from allocating the resource to $n$ individuals with identity $\theta$, and the utility function obtained by summing the score utilities across all individuals and diversity utility across all identities represents the preferences over sets of individuals.

Many widely used choice rules can be mapped to this framework. For example, a quota policy that restricts admission of individuals of each type $\theta$ by some $k_{\theta} \geq 0$ \citep{kojima2012school} can be rationalized by any strictly increasing $u$  and $\{h_{\theta}\}_{\theta \in \Theta}$ given by
    \begin{equation*}
        h_{\bm{\theta}}({N_{\theta}}(I)) = \begin{cases}
        0 &\text{ if } N_{\theta}(I) \leq k_{\theta} \\
        - q \bar u & \text{ if } N_{\theta}(I) > k_{\theta}
        \end{cases}
    \end{equation*}
     where $\bar u > \hat u \equiv \max_{s,s' \in \mathcal S} u(s) - u(s')$. Thus, a quota policy is rationalized by preferences where failing to meet the quota, which costs $q \bar u$, can never be remedied by improvements in score domain, which are capped by $q \hat u$.
     
     Another example is a reserve policy (see \cite{hafalir2013effective} and \cite{dur2020explicit}), which reserves $r_{\theta}$ of positions for individuals with each identity. These reserves and remaining open positions are then processed according to some order where in each step, highest scoring eligible individuals are chosen. When the number of individuals with each identity is higher than the number of reserve positions for that identity, and open positions are processed after all reserve positions, reserve policies can be rationalized by any strictly increasing $u$ and $\{h_{\theta}\}_{\theta \in \Theta}$ given by
    \begin{equation*}
        h_{{\theta}}({N_{\theta}}(I)) = \begin{cases}
        N_{\theta}(I) \bar u &\text{ if } N_{\theta}(I) \leq k_{\theta} \\
         k_{\theta} \bar u & \text{ if } N_{\theta}(I) > k_{\theta}
        \end{cases}
    \end{equation*}
    for $\bar u > \hat u$. This indicates that the diversity utility is increasing and more important than any gains in the score dimension until the reserve is met and is constant after the reserve requirements are satisfied. However, as the following example shows, if open positions are processed before reserves, the choice rule may fail acyclicity and cannot be represented by an additively separable utility function.
    \begin{example}
        There are two groups in one dimension, $\Theta=\{a,b\}$ and $q=3$. There is one reserve position for each of the groups. The processing order is open positions, group $a$ reserve and group $b$ reserve. Let $I = \{a_3,b_2,a_1,b_1\}$ and $I' = \{a_2,b_3,a_1,b_1\}$, where letters denote groups and subscripts denote scores.
         Both under $I$ and $I'$, the open position goes to the individual with score $3$ and the other individual from that group receives the reserve position: $C(I)=\{a_3,b_2,a_1\}$, implying $(1,a,2)>_C(1,b,2)$ and $C(I')=\{a_2,b_3,b_1\}$, implying $(1,b,2)>_C(1,a,2)$. This violates acyclicty and shows that this choice rule cannot be rationalized by a separable utility function.
\end{example}

The following proposition shows that this example can be generalized to any choice function that processes an open position before a reserve position.

\begin{prop}\label{prop:openfirst}
    Suppose that $C$ processes an open position before a reserve position. Then $C$ does not satisfy acyclicity, and thus is not rationalizable by additively separable preferences.
\end{prop}

The processing order of the positions is important for the distribution of positions to individuals from different groups. For example, \cite{dur2020explicit} shows that processing reserve positions of a group earlier is advantageous for that group and may be used as an additional lever in affirmative action programs. Proposition \ref{prop:openfirst} shows that if an institution adopts a processing order that processes open positions before reserve positions, this indicates a fundamental difference in preferences compared to the case where open positions are processed last.

\section{Conclusion}\label{sec:conclusion}

This paper contributes to the study of affirmative action and diversity concerns in market design. On the theoretical side, I introduce a model of diversity preferences and establish a connection between consumer theory and market design. I characterize when choice rules can be rationalized by diversity preferences, when they satisfy the substitutes condition, and when they can be rationalized by a utility function that treats diversity and quality separately. On the applied side, I show that the mechanism used to match workers to government jobs in India cannot be rationalized by diversity preferences and that considering intersectionality is crucial for the choice rule to satisfy the substitutes condition. Moreover, I identify the preferences that induce some well-known choice rules such as quotas and reserves. My framework provides a systematic way of evaluating the relationship between diversity preferences and the affirmative action policies implemented by the institutions that can be applied to many different real world markets.

\bibliographystyle{jfe}
\bibliography{bib-matching}

\begin{thebibliography}{41}
\expandafter\ifx\csname natexlab\endcsname\relax\def\natexlab#1{#1}\fi

\bibitem[{Abdulkadiro{\u{g}}lu(2005)}]{abdulkadirouglu2005college}
Abdulkadiro{\u{g}}lu, A., 2005. College admissions with affirmative action. International Journal of Game Theory 33, 535--549.

\bibitem[{Abdulkadiro{\u{g}}lu and S{\"o}nmez(2003)}]{abdulkadirouglu2003school}
Abdulkadiro{\u{g}}lu, A., S{\"o}nmez, T., 2003. School choice: A mechanism design approach. American economic review 93, 729--747.

\bibitem[{Adams(1965)}]{adams1965elements}
Adams, E.~W., 1965. Elements of a theory of inexact measurement. Philosophy of science 32, 205--228.

\bibitem[{Akbarpour et~al.(2021)Akbarpour, Budish, Dworczak, and Kominers}]{akbarpour2021economic}
Akbarpour, M., Budish, E.~B., Dworczak, P., Kominers, S.~D., 2021. An economic framework for vaccine prioritization. Available at SSRN 3846931 .

\bibitem[{Aygun and B{\'o}(2021)}]{aygun2021college}
Aygun, O., B{\'o}, I., 2021. College admission with multidimensional privileges: The brazilian affirmative action case. American Economic Journal: Microeconomics 13, 1--28.

\bibitem[{Ayg{\"u}n and S{\"o}nmez(2013)}]{aygun2013matching}
Ayg{\"u}n, O., S{\"o}nmez, T., 2013. Matching with contracts: Comment. American Economic Review 103, 2050--51.

\bibitem[{Ayg{\"u}n and Turhan(2017)}]{aygun2017large}
Ayg{\"u}n, O., Turhan, B., 2017. Large-scale affirmative action in school choice: Admissions to iits in india. American Economic Review 107, 210--13.

\bibitem[{Carvalho et~al.(2022)Carvalho, Pradelski, and Williams}]{carvalho2022affirmative}
Carvalho, J.-P., Pradelski, B., Williams, C., 2022. Affirmative action with multidimensional identities. Available at SSRN 4070930 .

\bibitem[{{\c{C}}elebi and Flynn(2022)}]{ccelebi2022priority}
{\c{C}}elebi, O., Flynn, J.~P., 2022. Priority design in centralized matching markets. The Review of Economic Studies 89, 1245--1277.

\bibitem[{{\c{C}}elebi and Flynn(2023)}]{celebi2021adaptive}
{\c{C}}elebi, O., Flynn, J.~P., 2023. Adaptive priority mechanisms. Working Paper .

\bibitem[{Chan and Eyster(2003)}]{chan2003does}
Chan, J., Eyster, E., 2003. Does banning affirmative action lower college student quality? American Economic Review 93, 858--872.

\bibitem[{Crenshaw(2013)}]{crenshaw2013demarginalizing}
Crenshaw, K., 2013. Demarginalizing the intersection of race and sex: A black feminist critique of antidiscrimination doctrine, feminist theory and antiracist politics. In: {\em Feminist Legal Theories\/}, Routledge, pp. 23--51.

\bibitem[{Dur et~al.(2018)Dur, Kominers, Pathak, and S{\"o}nmez}]{dur2018reserve}
Dur, U., Kominers, S.~D., Pathak, P.~A., S{\"o}nmez, T., 2018. Reserve design: Unintended consequences and the demise of boston’s walk zones. Journal of Political Economy 126, 2457--2479.

\bibitem[{Dur et~al.(2020)Dur, Pathak, and S{\"o}nmez}]{dur2020explicit}
Dur, U., Pathak, P.~A., S{\"o}nmez, T., 2020. Explicit vs. statistical targeting in affirmative action: Theory and evidence from chicago's exam schools. Journal of Economic Theory 187.

\bibitem[{Echenique and Yenmez(2015)}]{echenique2015control}
Echenique, F., Yenmez, M.~B., 2015. How to control controlled school choice. American Economic Review 105, 2679--94.

\bibitem[{Ehlers et~al.(2014)Ehlers, Hafalir, Yenmez, and Yildirim}]{ehlers2014school}
Ehlers, L., Hafalir, I.~E., Yenmez, M.~B., Yildirim, M.~A., 2014. School choice with controlled choice constraints: Hard bounds versus soft bounds. Journal of Economic theory 153, 648--683.

\bibitem[{Ellison and Pathak(2021)}]{ellison2021efficiency}
Ellison, G., Pathak, P.~A., 2021. The efficiency of race-neutral alternatives to race-based affirmative action: Evidence from chicago's exam schools. American Economic Review 111, 943--75.

\bibitem[{Fishburn(1970)}]{fishburn1970utility}
Fishburn, P.~C., 1970. Utility theory for decision making. Tech. rep., Research analysis corp McLean VA.

\bibitem[{Goto et~al.(2017)Goto, Kojima, Kurata, Tamura, and Yokoo}]{goto2017designing}
Goto, M., Kojima, F., Kurata, R., Tamura, A., Yokoo, M., 2017. Designing matching mechanisms under general distributional constraints. American Economic Journal: Microeconomics 9, 226--62.

\bibitem[{Grigoryan(2021)}]{grigoryan2021effective}
Grigoryan, A., 2021. Effective, fair and equitable pandemic rationing. Available at SSRN 3646539 .

\bibitem[{Hafalir et~al.(2013)Hafalir, Yenmez, and Yildirim}]{hafalir2013effective}
Hafalir, I.~E., Yenmez, M.~B., Yildirim, M.~A., 2013. Effective affirmative action in school choice. Theoretical Economics 8, 325--363.

\bibitem[{Hatfield and Kojima(2010)}]{hatfield2010substitutes}
Hatfield, J.~W., Kojima, F., 2010. Substitutes and stability for matching with contracts. Journal of Economic theory 145, 1704--1723.

\bibitem[{Hatfield and Milgrom(2005)}]{hatfield2005matching}
Hatfield, J.~W., Milgrom, P.~R., 2005. Matching with contracts. American Economic Review 95, 913--935.

\bibitem[{Hughes(2018)}]{hughes2018ethnic}
Hughes, M.~M., 2018. Ethnic quotas in electoral politics. Gender Parity and Multicultural Feminism: Towards a New Synthesis p.~97.

\bibitem[{Kamada and Kojima(2017)}]{kamada2017stability}
Kamada, Y., Kojima, F., 2017. Stability concepts in matching under distributional constraints. Journal of Economic theory 168, 107--142.

\bibitem[{Kamada and Kojima(2018)}]{kamada2018stability}
Kamada, Y., Kojima, F., 2018. Stability and strategy-proofness for matching with constraints: A necessary and sufficient condition. Theoretical Economics 13, 761--793.

\bibitem[{Kelso and Crawford(1982)}]{kelso1982job}
Kelso, A.~S., Crawford, V.~P., 1982. Job matching, coalition formation, and gross substitutes. Econometrica: Journal of the Econometric Society pp. 1483--1504.

\bibitem[{Kojima(2012)}]{kojima2012school}
Kojima, F., 2012. School choice: Impossibilities for affirmative action. Games and Economic Behavior 75, 685--693.

\bibitem[{Kojima et~al.(2020{\natexlab{a}})Kojima, Sun, and Yu}]{kojima2020job}
Kojima, F., Sun, N., Yu, N.~N., 2020{\natexlab{a}}. Job matching under constraints. American Economic Review 110, 2935--47.

\bibitem[{Kojima et~al.(2020{\natexlab{b}})Kojima, Sun, and Yu}]{kojima2020jobtax}
Kojima, F., Sun, N., Yu, N.~N., 2020{\natexlab{b}}. Job matching with subsidy and taxation. Available at SSRN 3624343 .

\bibitem[{Kurata et~al.(2017)Kurata, Hamada, Iwasaki, and Yokoo}]{kurata2017controlled}
Kurata, R., Hamada, N., Iwasaki, A., Yokoo, M., 2017. Controlled school choice with soft bounds and overlapping types. Journal of Artificial Intelligence Research 58, 153--184.

\bibitem[{Murota(1998)}]{murota1998discrete}
Murota, K., 1998. Discrete convex analysis. Mathematical Programming 83, 313--371.

\bibitem[{Murota et~al.(2016)}]{murota2016discrete}
Murota, K., et~al., 2016. Discrete convex analysis: A tool for economics and game theory. Journal of Mechanism and Institution Design 1, 151--273.

\bibitem[{Nishimura et~al.(2016)Nishimura, Ok, and Quah}]{nishimura2016}
Nishimura, H., Ok, E.~A., Quah, J. K.-H., 2016. {A Comprehensive Approach to Revealed Preference Theory}. Working Papers 201614, University of California at Riverside, Department of Economics.

\bibitem[{Pathak et~al.(2021)Pathak, S{\"o}nmez, {\"U}nver, and Yenmez}]{pathak2021fair}
Pathak, P.~A., S{\"o}nmez, T., {\"U}nver, M.~U., Yenmez, M.~B., 2021. Fair allocation of vaccines, ventilators and antiviral treatments: leaving no ethical value behind in health care rationing. In: {\em Working Paper\/}, pp. 785--786.

\bibitem[{Richter(1966)}]{richter1966revealed}
Richter, M.~K., 1966. Revealed preference theory. Econometrica: Journal of the Econometric Society pp. 635--645.

\bibitem[{Roth(1984)}]{roth1984stability}
Roth, A.~E., 1984. Stability and polarization of interests in job matching. Econometrica: Journal of the Econometric Society pp. 47--57.

\bibitem[{Roth(1985)}]{roth1985college}
Roth, A.~E., 1985. The college admissions problem is not equivalent to the marriage problem. Journal of economic Theory 36, 277--288.

\bibitem[{Scott(1964)}]{scott1964measurement}
Scott, D., 1964. Measurement structures and linear inequalities. Journal of mathematical psychology 1, 233--247.

\bibitem[{S{\"o}nmez and Yenmez(2022)}]{sonmez2022affirmative}
S{\"o}nmez, T., Yenmez, M.~B., 2022. Affirmative action in india via vertical, horizontal, and overlapping reservations. Econometrica 90, 1143--1176.

\bibitem[{Tversky(1964)}]{tversky1964optimal}
Tversky, A., 1964. Finite additive structures. Michigan Mathematical Psychology Program .

\end{thebibliography}

\begin{appendices}
\section{Proofs}\label{sec:proofs}
\subsection{Proof of Proposition \ref{prop:rationality}}
Define $>_{\mathcal S}$ as the empty relation. The result follows then from Proposition \ref{prop:monotonicity}.

\subsection{Proof of Proposition \ref{prop:increasingresponsiveness}}
    Suppose that $\succeq$ is increasing in scores. For any $i$ and $i'$ with  $\theta(i) = \theta(i')$ and $s(i) > s(i')$, let $h$ denote the function that maps $i$ to $i'$ and $i'$ to $i$, and is the identity function otherwise. Then for all $I$ with $\{i,i'\} \cap I = \emptyset$, $I \cup \{i\} >_{\mathcal S} I \cup \{i'\}$, which implies  $I \cup \{i\}\succ I \cup \{i'\}$.

    Conversely, suppose that $\succeq$ satisfies within-group responsiveness and $I >_{\mathcal S} I'$. Given the bijection $h$, let $I = \{i_1,\ldots,i_t\}$ and $I' = \{i_1',\ldots,i_t'\}$ where $i_j' = h(i_j)$. Let $I^j = \{i_1,\ldots,i_j,i_{j+1}',\ldots,i_t'\}$. Note that $I^t = I$ while $I^1 = I'$. Without loss of generality, let $s(i_1) > s(i_1')$ (such an $i$ exists as $I >_{\mathcal S} I'$). Moreover, $I = I^t \succeq I^{t-1} \succeq \ldots \succeq I^2 \succ I^1 = I'$, where at each $k$, $\succeq$ follows from within-group responsiveness (if $s(i_k) > s(i_k')$) or the fact that $\tau(I^k) = \tau(I^{k-1})$ (if $s(i_k) = s(i_k')$), yielding the result.

\subsection{Proof of Proposition \ref{prop:monotonicity}}
Define the relation $>_C$ as follows: $I >_C I'$ if $I \in C(\hat I)$ and $I' \subset \hat I$. Let $>_C \cup >_{\mathcal S}$ denote the union of $>_C$ and $>_{\mathcal S}$ and  $\text{tran}(>_C \cup >_{\mathcal S})$ denote the transitive closure of this relation. $C$ satisfies $>_{\mathcal S}$-congruence if (i) $I \text{ tran}(>_C \cup >_{\mathcal S}) I'$ and $I' \in C(\hat I)$ imply $I \in C(\hat I)$ for every $\hat I$ that contains $I$ and (ii)  $I \text{ tran}(>_C \cup >_{\mathcal S}) I'$ imply not $I'>_{\mathcal S} I$. The following lemma follows from the finiteness of $\mathcal I$.

\begin{lem}
$C$ satisfies $>_{\mathcal S}$-congruence if and only if $C$ doesn't admit a score-choice cycle.
\end{lem}
\begin{proof}
Suppose that $C$ admits score-choice cycle $I_1,\ldots,I_n$. Then for each $i \leq n-1$, either $I_i >_C I_{i+1}$ or $I_i >_{\mathcal S} I_{i+1}$. Thus,  $I_1 \text{ tran}(>_C \cup >_{\mathcal S}) I_n$. Moreover, as either $I_n >_{\mathcal S} I_1$ or $I_n \in C(\hat I)$, $I_n \not \in C(\hat I)$ and $I_1 \subset \hat I$ for some $\hat I$, $>_{\mathcal S}$-congruence fails.

Conversely, suppose that $C$ doesn't satisfy  $>_{\mathcal S}$-congruence and let $I_1,\ldots,I_n$ denote the sets that cause the violation. Then $I_1 \text{ tran}(>_C \cup >_{\mathcal S}) I_n$. Then for each $i<n$, either (i) there exists an $\hat I_i$ such that $I_i \in C(\hat I_i)$ and $I_{i+1} \subset \hat I_i$ or (ii) $I_i >_{\mathcal S} I_{i+1}$. Moreover, we also have either (i) there exists $\hat I_n$ such that $I_n \in C(\hat I_n)$, $I_{1} \subset \hat I_n$ and $I_{1} \not \in C(\hat I_n)$ or (ii) $I_{n} >_{\mathcal S} I_{1}$, which completes the proof.
\end{proof}
The result then follows from Theorem 7 in \cite{nishimura2016}, who shows that a choice rule $C$ is rationalizable by a preference relation $\succeq$ that extends $>_{\mathcal S}$ (which is equivalent to $\succeq$ is increasing in scores) if and only if it satisfies $>_{\mathcal S}$-congruence.

\subsection{Proof of Proposition \ref{incompatibility:prop}}

Suppose that $\mathcal J$ includes $q$ individuals from each $\theta \in  \Theta$ and $\succeq$ doesn't consider intersectionality. I say that $i$ is a $(j,k)$ individual if $\theta_j(i) = k$.

Let $D^*$ denote the set of all optimal marginal distributions. Formally, $d \in  D^*$ if there exists $I'$ such that $I' \in C(\mathcal J)$ and $   M(I') = d$. Let $d_1$ denote an element of $D^*$ with the highest number of $(1,1)$ individuals. $m_{11}$ denotes the number of $(1,1)$ individuals at $d_1$. $D^*_1$ denotes the set of all optimal group distributions where the number of $(1,1)$ individuals is $m_{11}$. Let $d^*_{11}$ be a group distribution in $D^*_1$ with the highest number of $(2,1)$ individuals.  $m_{21}$ denotes the number of $(2,1)$ individuals at $d^*_{11}$.

A set of individuals $I$ is \textit{compatible with} marginal distributions $d^*$ if there exists $I'$ such that $M(I \cup I') = d^*$ and $I'$ is a \textit{complement} of $I$ for $d^*$. Let $M_{ij}(I)$ ($M_{ij}(d)$) denote the number of group $i$ individuals in dimension $j$ in $I$ ($d$).

\begin{lem}\label{lem:compatibility}
If $M_{ij}(I) \leq M_{ij}(d)$ for all $i$ and $j$, then $I$ is compatible with $d$.
\end{lem}
\begin{proof}
If $M_{ij}(I) < M_{ij}(d)$ for some $ij$, then for each dimension $i'$, there exists a group $j'$ such that $M_{i'j'}(I) < M_{i'j'}(d)$. Let $t$ denote an individual who belongs to group $j'$ at each dimension $i'$. Then the set $I \cup \{t\}$ still satisfies $M_{ij}(I) \leq M_{ij}(d)$ and repeating this procedure yields a $\tilde I$ such that $M_{ij}(\tilde I) = M_{ij}(d)$ and $I$ is compatible with $d$.
\end{proof}

\textbf{Case 1: $m_{11} \leq m_{21}$.}
\begin{clm}
There exists $I_{11} = \{i_1^{11},\ldots,i_{m_{11}}^{11}\}$ compatible with $d^*_{11}$ where all $i \in I_{11}$ are $(1,1)$ and $(2,1)$ individuals.
\end{clm}
\begin{proof}
Since groups $(1,1)$ and $(2,1)$ have (weakly) more individuals at $d^*_{11}$, one can choose the groups of individuals in $I_{11}$ in other dimensions to satisfy $M_{ij}(I_{11}) \leq M_{ij}(d_{11}^*)$ for all $i$ and $j$. Then the result follows from Lemma \ref{lem:compatibility}.
\end{proof}

Let $I'$ denote a complement of $I_{11}$ at $d^*_{11}$. Take $j \in I_{11}$ and $k \in I'$, where $k$ isn't a $(1,1)$ or $(2,1)$ individual.\footnote{This is possible since the preferences value diversity and all individuals in $I_{11}$ are both $(1,1)$ and $(2,1)$ indviduals.} Define $\tilde j$ and $\tilde k$ as

\begin{equation*}
    \theta_1(\tilde j) = \theta_1(j),  \theta_{\ell}(\tilde i) = \theta_{\ell}(k) \text{ for all } \ell \neq 1
\end{equation*}
\begin{equation*}
    \theta_1(\tilde k)=\theta_1(k),\theta_{\ell}(\tilde k)=\theta_{\ell}(j) \text{ for all } \ell \neq 1
\end{equation*}

Let $\tilde I_{11} = I_{11} \setminus \{j\} \cup \tilde j$ and $I'' = I' \setminus \{k\} \cup \tilde k$. Note that $I''$ is a complement of $\tilde I_{11}$ at $d^*_{11}$. The following claim holds by construction.

\begin{clm}
$I'$ and $I''$ doesn't have any $(1,1)$ individuals. Moreover, $I''$ has $m_{21}-m_{11} + 1$ group $(2,1)$ individuals.
\end{clm}

Let $\bar I = I_{11} \cup I' \cup \{\tilde j, \tilde k\}$. As  $M(I_{11} \cup I') = d^*_{11}$,  $I_{11} \cup I' \in C(\bar I)$.

\begin{lem}\label{case1lemma}
There doesn't exist an $I^* \in C(\bar I \setminus \{k\})$ such that $I_{11} \subset I^*$
\end{lem}
\begin{proof}
Suppose that there is such an $I^*$. Then it must be that, $\tilde j \not \in I^*$, since otherwise there will be more than $m_{11}$ $(1,1)$ individuals at  $I^*$, which is a contradiction. However, this means that  $I^* = I_{11} \cup I''$. But then $I^*$ has $m_{21}+1$ $(2,1)$ individuals and $m_{11}$ $(1,1)$ individuals, which contradicts the optimality of $I^*$ as  $d^*_{11}$ is a group distribution in $D^*_1$ with the highest number of $(2,1)$ individuals and has $m_{21}$ such individuals. Since $\tilde I_{11} \cup I''$ is available and optimal, this is a contradiction.
\end{proof}
The result then follows from the fact that $I_{11}$ is chosen from $\bar I$, but not from $\bar I \setminus \{k\}$.\\

\textbf{Case 2: $m_{11} > m_{21}$}. Let $n = m_{11}-m_{21}$. 

\begin{clm}
There exists $I_{12} = \{i_1^{11},\ldots,i_{m_{21}}^{11},i_1^{1},\ldots,i_n^{1}\}$ where the first $m_{21}$ elements are $(1,1)$ and $(2,1)$ individuals, rest are $(1,1)$ individuals and $I_{12}$ is compatible with $d^*_{11}$.
\end{clm}
\begin{proof}
Since groups $(1,1)$ and $(2,1)$ have (weakly) more individuals at $d^*_{11}$, one can choose the groups of individuals in $I_{12}$ to satisfy $M_{ij}(I_{12}) \leq M_{ij}(d_{11}^*)$ for all $i$ and $j$. Then the result follows from Lemma \ref{lem:compatibility}.
\end{proof}

Let $I'$ denote a complement of $I_{12}$ at $d^*_{11}$. Take $j \in I_{12}$ and $k \in I'$, where $k$ isn't a $(1,1)$ or $(2,1)$ individual.\footnote{This is possible since the preferences value diversity and all individuals in $I_{12}$ are both $(1,1)$ and $(2,1)$ individuals.}  Define $\tilde j$ and $\tilde k$ as

\begin{equation*}
    \theta_2(\tilde j) = \theta_2(j),  \theta_{\ell}(\tilde j) = \theta_{\ell}(k) \text{ for all } \ell \neq 1
\end{equation*}
\begin{equation*}
    \theta_2(\tilde k) = \theta_2(k),  \theta_{\ell}(\tilde k) = \theta_{\ell}(j) \text{ for all } \ell \neq 1
\end{equation*}

Let $\tilde I_{12} = I_{12} \setminus \{j\} \cup \tilde j$ and $I'' = I' \setminus \{k\} \cup \tilde k$. Note that $I''$ is a complement of $\tilde I_{12}$ at $d^*_{11}$.  The following claim holds by construction.

\begin{clm}
$I'$ and $I''$ doesn't have any $(2,1)$ individuals. Moreover, $I''$ has $1$ group $(1,1)$ individual.
\end{clm}

Let $\bar I = I_{12} \cup I' \cup \{\tilde j, \tilde k\}$. First, note that $I_{12} \cup I' \in C(\bar I)$ and $\tilde I_{12} \cup I'' \in C(\bar I)$, since  $M(I_{12} \cup I') = M(\tilde I_{12} \cup I'') = d^*_{11}$. 

\begin{lem}\label{case2lemma}
There doesn't exist an $I^* \in C(\bar I \setminus \{k\})$ such that $I_{12} \subset I^*$.
\end{lem}
\begin{proof}
Since $\tilde I_{12} \cup I''$ is available and optimal, $I^*$ must also be optimal.
For a contradiction, suppose that such an $I^*$ exists. Then it must be that, $\tilde j \not \in I^*$, as otherwise $I^*$ would have $m_{11}$ $(1,1)$ individuals and $m_{21}+1$ $(2,1)$ individuals.

However, this means that $I^* = I_{11} \cup I''$. But then $I^*$ has $m_{11}+1$ $(1,1)$ individuals, which contradicts the optimality of $I^*$.
\end{proof}
As $I_{11}$ is chosen from $\bar I$, but not from $\bar I \setminus \{k\}$, the result follows.

\subsection{Proof of Proposition \ref{GSrepresentation}}

Suppose that $C$ satisfies substitutes, within-group responsiveness and acyclicty. Observe that $(s,\theta,n) >_C (s',\theta',n')$ implies that $(s,\theta,n) \neq (s',\theta',n')$. Let $\mathcal S = \{s_0,\ldots,s_K\}$ denote the ordered set of scores. Let $H(s,\theta,n)$ denote a set formed by $n$ individuals of type $(s,\theta)$.
\begin{lem}\label{lem:monotone}
    For each $i > 0$, there exist $\theta$ and $n$ such that $(s_i,\theta,n) >_C (s_{i-1},\theta,n)$.
\end{lem}
\begin{proof}
    Take arbitrary $\theta \neq \theta'$. Consider $I = H(s_i,\theta,q) \cup  H(s_i,\theta',q) \cup  \{k_{\theta},k_{\theta'}\}$, where $t(k_{\theta}) = (s_{i-1},\theta)$ and $t(k_{\theta'}) = (s_{i-1},\theta')$. Suppose that $\hat I \in C(I)$. By within-group responsiveness, either $k_{\theta} \not \in \hat I$ or $k_{\theta'} \not \in \hat I$. Wlog, let $k_{\theta} \not \in \hat I$ and $n = N_{\theta}(I)$. Then  $(s_i,\theta,n) >_C (s_{i-1},\theta,n)$.
\end{proof}

For each $\theta$, we compute the number of $\theta$ individuals who are guaranteed to be chosen.
\begin{equation}
    I_{\theta} = \left( \bigcup_{\theta' \neq \theta,s \in \mathcal S} H(s,\theta',q) \right) \cup H(s_0,\theta,q)
\end{equation}

Take a $J \in C(I_{\theta})$ and let $n_{\theta} = N_{\theta}(J)$. 

\begin{lem}\label{lem:concavity}
    For each integer $n \in [n_{\theta},q-1]$, $(s_0,\theta,n)>_C(s_0,\theta,n+1)$.
\end{lem}
\begin{proof}
    Let $\hat I = H(s_0,\theta,q)$. As $J \in C(I_{\theta})$, by substitutes, $J \in C(J \cup \hat I)$. Remove $n-n_{\theta}$ non $\theta$ individuals from $J$ to define $\tilde J$. By substitutes, there exists $I' \in C(\tilde J \cup \hat I)$ such that all $i \in \tilde J$ and $\theta(i) \neq \theta$ are in $I'$. Thus, $N_{\theta}(I') = n$, proving $(s_0,\theta,n)>_C(s_0,\theta,n+1)$.
\end{proof}

\begin{lem}
    Suppose that $I^* \in C(I)$ and $N_{\theta}(I^*) < n_{\theta}$. Then all group $\theta$ agents in $I$ are in $I^*$.
\end{lem}
\begin{proof}
    Suppose that this doesn't hold. Then there exists $I$ and $i \in I$ such that $I^* \in C(I)$, $\hat n = N_{\theta}(I^*) < n_{\theta}$, $\theta(i) = \theta$ and $i \not \in I^*$. By substites, $I^* \in C(I^* \cup \{i\})$. Let $\hat I$ denote the set of non-$\theta$ individuals in $I^*$. Let $\tilde I = H(s_0,\theta,q)$.
    \begin{clm}
        $I^* \in C(I^* \cup \{i\} \cup \tilde I)$
    \end{clm}
    \begin{proof}
        If $I^* \not \in C(I^* \cup \{i\} \cup \tilde I)$, then there is $I' \in C(I^* \cup \{i\} \cup \tilde I)$ such that $N_{\theta}(I) > \hat n$ and by within-group responsiveness $I$ includes all identity $\theta$ individuals in $I^* \cup \{i\}$. Then by substitutes, there exists $I_1 \in C(I^* \cup \{i\})$ such that $N_{\theta}(I_1) = \hat n +1$. Then $\tau(I^*) \neq \tau(I_1)$, which is a contradiction as both are chosen from $I^* \cup \{i\}$.
    \end{proof}
    As $I^* \in C(I^* \cup \tilde I \cup \{i\})$, by substitutes, $I^* \in C(I^* \cup \tilde I)$. As $\hat I \subseteq I^*$, there is $I_2 \in C(\hat I \cup \tilde I)$ such that $\hat I \subseteq I_2$, which implies that $N_{\theta}(I_2) = \hat n < n_{\theta}$. By construction, there exists $I_{\theta}' \subset I_{\theta}$ such that $\tau(I_{\theta}') = \tau(\hat I \cup \tilde I)$. Moreover, by substitutes, there is $J' \in C(I_{\theta}')$ such that $N_{\theta}(J') = n_{\theta}$. However, this is a contradiction as $\tau(J') \neq \tau(I_2$), $J' \in C(I_{\theta}')$ and $I_2 \in C(\hat I \cup \tilde I)$.
    \end{proof}

\begin{lem}
There exist $u$ and $h$ such that $s,d >_C s',d'$ implies $u(s)+h(d) > u(s') + h(d')$.
\end{lem}
\begin{proof}
Follows from Theorem 4.1 in \cite{fishburn1970utility}.
\end{proof}

By Lemma \ref{lem:monotone} $u$ is strictly increasing. To ensure concavity, I will modify $h$ without changing the sets of chosen individuals. Let $\bar u = \max_{\theta,n} h(\theta,n) + u(s_k)$ as the largest utility contribution an individual can have under $u$ and $h$. Define $\tilde h$ as follows:

\begin{equation}
    \tilde h(\theta,n) = \begin{cases}
        h(\theta,n) & \text{ if } n > n_{\theta}\\
        \bar u & \text{ if } n \leq n_{\theta}
    \end{cases}
\end{equation}

Let $\tilde h_{\theta}(n) = \sum_{i=1}^{n} \tilde h(\theta,n)$. By Lemma \ref{lem:concavity}, $\tilde h_{\theta}$ is concave.

\begin{lem}
$U(I)$ where
    \begin{equation*}
        U(I) = \sum_{i \in I} u(s(i)) + \sum_{\theta \in \Theta} \tilde h_{\theta}(N_{\theta}(I))
    \end{equation*}
rationalizes $C$.    
\end{lem}
\begin{proof}
For a contradiction, assume it doesn't rationalize $C$. Then there exists $q-$element subsets $I$ and $I'$ such that $U(I) > U(I')$,  $I' \in C(\hat I)$ for some $\hat I$ that includes $I$. Moreover, we can take $I$ to be a maximizer of $U(\tau(I)) = \max_{\tilde I \in 2_q^{\hat I}} U(\tau(\tilde I))$, which exists by the finiteness of $\hat I$.

First, if there exists $i \in I \setminus I'$ and $j \in I' \setminus I$ such that $t(i)=t(j)$, let $\tilde I = I' \setminus \{j\} \cup \{i\}$. 
Note that the statement $\tau(I) \neq \tau(\tilde I)$, $U(I) > U(\tilde I)$,  $\tilde I \in C(\hat I)$ for some $\hat I$ that includes $I$ still holds.
We can repeat this until there doesn't exist any $i \in I \setminus \tilde I$ and $j \in \tilde I \setminus I$ such that $t(i)=t(j)$.

Choose an arbitrary $i \in \tilde I \setminus  I$. Since $C$ satisfies substitutes, there exists $I_C$ such that $I_C \in C(I \cup \{i\})$ and $i \in I_C$. Thus, there exists $j \in I$ such that $j \not \in I_C$. As $t(i) \neq t(j)$, if $\theta(i) = \theta(j)$, within-group responsiveness of $C$  implies $s(i) > s(j)$, which implies $U(I \setminus \{i\} \cup \{j\}) > U(I)$, which contradicts that $I$ is a maximizer of $U$. Thus, $\theta(i) \neq \theta(j)$, which implies $I_C \setminus \{i\} \cup \{j\} \not \in C(I \cup \{i\})$, which implies $(s(i),\theta(i),N_{\theta(i)}(I_C)) >_C (s(j),\theta(j),N_{\theta(j)}(I_C)+1)$. Then $$u(s(i)) + h(\theta(i),N_{\theta(i)}(I_C)) > u(s(j)) + h(\theta(j),N_{\theta(j)}(I_C)+1)$$

Moreover, as $j \not \in I_C$, $N_{\theta(j)}(I_C)+1 > n_{\theta(j)}$, and therefore
$$u(s(i)) + \tilde h(\theta(i),N_{\theta(i)}(I_C)) > u(s(j)) + \tilde h(\theta(j),N_{\theta(j)}(I_C)+1)$$
However, above equation indicates $U(I \cup \{i\} \setminus \{j\}) > U(I)$, which is a contradiction as $I$ maximizes utility in $\hat I$, which includes $I \cup \{i\} \setminus \{j\}$.
\end{proof}

To prove the second part let $C$ denote the indued choice function. Given $h_{\theta}(n)$, define $ h(\theta,n) = h_{\theta}(n) - h_{\theta}(n-1)$, which are concave in $n$ as $h_{\theta}(n)$ are concave. Assume for a contradiction there exists a cycle at $>_C$.
This means that for each $(s_i,d_i)$ and $(s_i',d_i')$, $u(s_i)+h(d_i) > u(s_i')+h(d_i')$, which implies $\sum_{i}(s_i',d_i')$, $u(s_i)+h(d_i) > \sum_{i} u(s_i')+h(d_i')$, which is a contradiction $(s_i',d_i')$ is a permutation of $(s_i,d_i)$. Within-group responsiveness of $C$ is immediate as $u$ is strictly increasing.

To show that $C_U$ satisfies substitutes, suppose that $I'_1 \subseteq I_1 \in C(\hat I_1)$ and  $\hat I_2 \subseteq \hat I_1$. Take any $I_2' \subseteq I_1' \cap \hat I_2$. I will show that there exists $\tilde I \in C(\hat I_2)$ such that $I_2' \subseteq \tilde I$. Take $I_2 \in C(\hat I_2)$ and suppose that $i \in I'_2$ but $i \not \in I_2$.

\begin{clm}\label{clm:substitutes}
    There exists $j \in I_2$, $j \not \in I'_2$ and $I_2 \setminus \{j\} \cup \{i\} \in C(\hat I_2)$.
\end{clm}
\begin{proof}
    Let $\theta(i) = \hat \theta$. First, if $N_{\hat \theta}(I_1) \leq N_{\hat \theta}(I_2)$, then there exists $j \in I_2$, $j \not \in I_1$ $\theta(j) = \hat \theta$. Moreover, as $i \in I_2' \subseteq I_1 \in C(\hat I_1)$ and $j \not \in I_1$, $s(i) \geq s(j)$, as otherwise this would be a contradiction that $u$ is increasing in $s$. Then $U(I_2 \setminus \{j\} \cup \{i\}) \geq U(I_2)$, which proves the result.

    Second, if $N_{\hat \theta}(I_1) > N_{\hat \theta}(I_2)$, then there exists $j \in I_2$, $j \not \in I_1$, $\theta(j) = \theta' \neq \hat \theta$ such that $N_{\theta'}(I_2) > N_{\theta'}(I_1)$. As $j \not \in I_1$
    \begin{equation}
        s(i) + h(\hat \theta,N_{\hat \theta}(I_1)) \geq s(j) + h(\theta',N_{\theta'}(I_1))
    \end{equation}
    As $h_{\theta}$ are concave for all $\theta$, we have     
    \begin{equation}
        s(i) + h(\hat \theta,N_{\hat \theta}(I_2)) \geq s(j) + h(\theta',N_{\theta'}(I_2))
    \end{equation}
    which implies that $U(I_2 \setminus \{j\} \cup \{i\}) \geq U(I_2)$ and proves the result.
\end{proof}

Repeatedly applying Claim \ref{clm:substitutes}, starting with any $I_2 \in C(\hat I_2)$, we arrive at a $\tilde I \in C(\hat I_2)$ such that $I'_2 \subseteq \tilde I$, which shows that $C$ satisfies substitutes.

\subsection{Proof of Proposition \ref{prop:openfirst}}

Let $\theta_1$ denote a group with a reserve position that is processed after the final open position that preceeds reserve positions. Let $I$ denote a set of individuals that have $q$ individuals from $\theta_1$ and $\theta_2$, where all $\theta_1$ individuals have scores $s_0$ and all $\theta_2$ individuals have scores $s_{K-1}$ and for all $j \geq 3$, $r_j$ individuals from $\theta_j$ with scores $s_1$. Let $r_1$ and $r_2$ denote the number of reserve positions for $\theta_1$ and $\theta_2$ and $o$ denote the number of open positions. Under $I$ all $r_j$ reserve positions are assigned to $\theta_j$ individuals, while open positions are assigned to $\theta_2$ individuals, giving $(s_{K_1},\theta_2,r_2+o) >_C (s_0,\theta_1,r_1+1)$. 

Let $n$ denote the number of $\theta_1$ reserve positions before the first open position. Define $\hat I$ by increasing the scores of $n+1$ $\theta_1$ individuals to $s_K$. At $\hat I$, one open position is assigned to a $\theta_1$ individual, while all other open positions are assigned to $\theta_2$ individuals. Thus there are $r_1+1$ $\theta_1$ such individuals and $r_2+o-1$ $\theta_2$ individuals in $C(\hat I)$. Thus, $(s_0,\theta_1,r_1+1) >_C (s_{K_1},\theta_2,r_2+o)$, violating acyclicity.

\section{Extensions}

\subsection{Utility Representation}\label{app:utility}
In this section, I state the results in Section \ref{sec:rationality} for a utility function $U$ instead of a preference relation $\succeq$. The preferences of the institution are represented by a function $U: \mathcal T \to \mathbb R$ and I will write $U(I)$ instead of $U(\tau(I))$. $U$ rationalizes $C$ if $C$ always chooses the $U$-maximal sets of individuals, that is, for all $|I| \geq q$, $C(I) = \{I': U(\tau(I')) = \max_{\hat I \in 2_q^{I}} U(\tau(\hat I))\}$. $U$ is increasing in scores if  $I  >_{\mathcal S} I'$ implies $U(I) > U(I')$. Both propositions are immediate consequences of Propositions \ref{prop:rationality} and \ref{prop:monotonicity} and finiteness of $\mathcal T$.

\begin{prop}
$C$ doesn't admit a choice cycle if and only if there exists a $U$ that rationalizes $C$.
\end{prop}

\begin{prop}
$C$ doesn't admit a score-choice cycle if and only if there exists a preference relation $\succeq$ that rationalizes $C$ and is increasing in $\mathcal S$.
\end{prop}

\subsection{Rationalizability of the Supreme Court Mandated Choice Rule}\label{sec:sciwithoutscores}

The next example shows that the shortcomings of the supreme court mandated choice rule are even greater, and it cannot be rationalized at all.

\begin{example}\label{ex:SCInonmonotone}
$I = \{m_1^g,m_2^g,w_1^r,w_2^r,w_1^g\}$. Capacity and reserves are $o=2$, $r=1$, $o_w=1$ and  $r_w=0$. The scores of individuals are given by 
$$s(m_1^g) > s(m_2^g) > s(w_1^r) > s(w_2^r) > s(w_1^g)$$

As there are two open positions and $m_1^g$ and $m_2^g$ are the two highest scoring individuals, $\mathcal M = \emptyset$ and only $I^g$ are eligible for the open positions.  Thus, in the first stage, $m_1^g$ and $w_1^g$ are chosen. In the second stage, highest scoring reserve-eligible individual, $w_1^r$ is chosen. This means that the set $I_1 = \{m_1^g,w_1^r,w_1^g\}$ is chosen when $I_2 = \{m_1^g,w_1^r,w_2^r\}$ is available. 

Now consider $\tilde I = \{m_1^g,w_1^r,w_2^r,w_1^g\}$, which removes $m_2^g$ from the set of applicants. Then $\mathcal M = \{w_1^r\}$ and in the first stage, $m_1^g$ and $w_1^r$ are chosen. In the second stage, the only remaining reserve-eligible individual, $w_2^r$ is chosen. Therefore, $I_2$ is chosen when $I_1$ is available and $I_1,I_2,I_1$ is a choice cycle.
\end{example}

This example shows that the $C_S$ rule doesn't only violate within-group responsiveness, but it also cannot be result of preferences of a rational decision-maker. 

\subsection{Heterogeneous Qualities and Gross Substitutes}\label{subsec:hetero}
This section extends the analysis to the setting where $|\mathcal S| > 1$. When $\succeq$ is increasing in scores (or equivalently, satisfies within-group responsiveness), the scores in this model are analogous to (inverse) salaries in  \cite{kelso1982job}, where a higher salary is worse for the institution. Therefore, I adopt the following gross substitutes definition given in \cite{kelso1982job}. I use $s(I)$ to denote the vector of scores of individuals in $I$.

\begin{defn}
Let $\tilde I \subseteq \hat I \in C(I)$. Define $I'$ by (weakly) decreasing the scores of all $I \setminus \tilde I$. If $C$ satisfies gross substitutes, then there exists $\bar I$ such that $\tilde I \subset \bar I$ and $\bar I \in C(I')$.
\end{defn}

Gross substitutes condition requires that if a set of individuals are chosen, and the scores of other individuals decrease, then that set of individuals must still be chosen. I also extend the definition of preferences that don't consider intersectionality to settings with heterogeneous qualities.

\begin{defn}
    $\succeq$ does not consider intersectionality if  $\{
     s(I), M(I)\} = \{ s(I'), M(\tilde I')\}$ implies $I \sim I'$.
\end{defn}

With heterogeneous qualities, an institution does not consider intersectionality is indifferent between two sets of individuals whenever they have the same cross-sectional distribution of groups and the same scores. The following proposition shows that the the relationship between intersectionality and the substitutes condition generalizes to this setting. 

\begin{prop}\label{incompatibilitywithscores:prop}
Suppose that $C_{\succeq}$ is induced by $\succeq$ that does not consider intersectionality, satisfies within-group responsiveness and values diversity. Then $C_{\succeq}$ doesn't satisfy gross substitutes.
\end{prop}

\begin{proof}
The proof closely follows the proof of Proposition \ref{incompatibility:prop} with minor modifications, and included for completeness.  Let $\mathcal S = \{s_0,\ldots,s_K\}$ denote the ordered set of scores.

Suppose that $\mathcal J$ includes $q$ individuals from each $\theta \in  \Theta$ with scores $s_K$  and $\succeq$ doesn't consider intersectionality. I say that $i$ is a $(j,k)$ individual if $\theta_j(i) = k$.

Let $D^*$ denote the set of all optimal marginal distributions when all individuals have maximum score $s_K$. Formally, $d \in  D^*$ if there exists $I'$ such that $I' \in C(\mathcal J)$ and $   M(I') = d$. Let $d_1$ denote an element of $D^*$ with the highest number of $(1,1)$ individuals. $m_{11}$ denotes the number of $(1,1)$ individuals at $d_1$. $D^*_1$ denotes the set of all optimal group distributions where the number of $(1,1)$ individuals is $m_{11}$. Let $d^*_{11}$ be a group distribution in $D^*_1$ with the highest number of $(2,1)$ individuals.  $m_{21}$ denotes the number of $(2,1)$ individuals at $d^*_{11}$.

A set of individuals $I$ is \textit{compatible with} marginal distributions $d^*$ if there exists $I'$ such that $M(I \cup I') = d^*$ and $I'$ is a \textit{complement} of $I$ for $d^*$. Let $M_{ij}(I)$ ($M_{ij}(d)$) denote the number of group $i$ individuals in dimension $j$ in $I$ ($d$).

\begin{lem}\label{lem:compatibilitygs}
If $M_{ij}(I) \leq M_{ij}(d)$ for all $i$ and $j$, then $I$ is compatible with $d$.
\end{lem}
\begin{proof}
If $M_{ij}(I) < M_{ij}(d)$ for some $ij$, then for each dimension $i'$, there exists a group $j'$ such that $M_{i'j'}(I) < M_{i'j'}(d)$. Let $t$ denote an individual who belongs to group $j'$ at each dimension $i'$. Then the set $I \cup \{t\}$ still satisfies $M_{ij}(I) \leq M_{ij}(d)$ and repeating this procedure yields a $\tilde I$ such that $M_{ij}(\tilde I) = M_{ij}(d)$ and $I$ is compatible with $d$.
\end{proof}

\textbf{Case 1: $m_{11} \leq m_{21}$.}
\begin{clm}
There exists $I_{11} = \{i_1^{11},\ldots,i_{m_{11}}^{11}\}$ compatible with $d^*_{11}$ where all $i \in I_{11}$ are $(1,1)$ and $(2,1)$ individuals.
\end{clm}
\begin{proof}
Since groups $(1,1)$ and $(2,1)$ have (weakly) more individuals at $d^*_{11}$, one can choose the groups of individuals in $I_{11}$ in other dimensions to satisfy $M_{ij}(I_{11}) \leq M_{ij}(d_{11}^*)$ for all $i$ and $j$. Then the result follows from Lemma \ref{lem:compatibilitygs}.
\end{proof}

Let $I'$ denote a complement of $I_{11}$ at $d^*_{11}$. Take $j \in I_{11}$ and $k \in I'$, where $k$ isn't a $(1,1)$ or $(2,1)$ individual.\footnote{This is possible since the preferences value diversity and all individuals in $I_{11}$ are both $(1,1)$ and $(2,1)$ indviduals.} Define $\tilde j$ and $\tilde k$ as

\begin{equation*}
    \theta_1(\tilde j) = \theta_1(j),  \theta_{\ell}(\tilde i) = \theta_{\ell}(k) \text{ for all } \ell \neq 1
\end{equation*}
\begin{equation*}
    \theta_1(\tilde k)=\theta_1(k),\theta_{\ell}(\tilde k)=\theta_{\ell}(j) \text{ for all } \ell \neq 1
\end{equation*}

Let $\tilde I_{11} = I_{11} \setminus \{j\} \cup \tilde j$ and $I'' = I' \setminus \{k\} \cup \tilde k$. Note that $I''$ is a complement of $\tilde I_{11}$ at $d^*_{11}$. The following claim holds by construction.

\begin{clm}
$I'$ and $I''$ doesn't have any $(1,1)$ individuals. Moreover, $I''$ has $m_{21}-m_{11} + 1$ group $(2,1)$ individuals.
\end{clm}

Let $\bar I = I_{11} \cup I' \cup \{\tilde j, \tilde k\}$. As  $M(I_{11} \cup I') = d^*_{11}$,  $I_{11} \cup I' \in C(\bar I)$. Moreover, let $\hat k$ denote an individual with $\theta(\hat k) = \theta(k)$ and $s(\hat k)< s_K$. Let $\bar I_{\hat k} = \bar I \cup \{\hat k\} \setminus \{k\}$.

\begin{lem}\label{case1lemmags}
There doesn't exist an $I^* \in C(\bar I_{\hat k})$ such that $I_{11} \subset I^*$
\end{lem}
\begin{proof}
Suppose that there is such an $I^*$. First, note that $\hat k \not \in I^*$, as $\tilde I_{11} \cup I'' \succ J$ for all $J$ that includes $\hat k$ by within-group responsiveness.

Moreover, it must be that, $\tilde j \not \in I^*$, since otherwise there will be more than $m_{11}$ $(1,1)$ individuals at  $I^*$, which is a contradiction. However, this means that either $I^* = I_{11} \cup I''$. But then, $I^*$ has $m_{21}+1$ $(2,1)$ individuals and $m_{11}$ $(1,1)$ individuals, which contradicts the optimality of $I^*$ as  $d^*_{11}$ is a group distribution in $D^*_1$ with the highest number of $(2,1)$ individuals and has $m_{21}$ such individuals. Since $\tilde I_{11} \cup I''$ is available and optimal, this is a contradiction.
\end{proof}
The result then follows from the fact that $I_{11}$ is chosen from $\bar I$, but not from $\bar I_{\hat k}$.\\

\textbf{Case 2: $m_{11} > m_{21}$}. Let $n = m_{11}-m_{21}$. 

\begin{clm}
There exists $I_{12} = \{i_1^{11},\ldots,i_{m_{21}}^{11},i_1^{1},\ldots,i_n^{1}\}$ where the first $m_{21}$ elements are $(1,1)$ and $(2,1)$ individuals, rest are $(1,1)$ individuals and $I_{12}$ is compatible with $d^*_{11}$.
\end{clm}
\begin{proof}
Since groups $(1,1)$ and $(2,1)$ have (weakly) more individuals at $d^*_{11}$, one can choose the groups of individuals in $I_{12}$ to satisfy $M_{ij}(I_{12}) \leq M_{ij}(d_{11}^*)$ for all $i$ and $j$. Then the result follows from Lemma \ref{lem:compatibilitygs}.
\end{proof}

Let $I'$ denote a complement of $I_{12}$ at $d^*_{11}$. Take $j \in I_{12}$ and $k \in I'$, where $k$ isn't a $(1,1)$ or $(2,1)$ individual.\footnote{This is possible since the preferences value diversity and all individuals in $I_{12}$ are both $(1,1)$ and $(2,1)$ individuals.}  Define $\tilde j$ and $\tilde k$ as

\begin{equation*}
    \theta_2(\tilde j) = \theta_2(j),  \theta_{\ell}(\tilde j) = \theta_{\ell}(k) \text{ for all } \ell \neq 1
\end{equation*}
\begin{equation*}
    \theta_2(\tilde k) = \theta_2(k),  \theta_{\ell}(\tilde k) = \theta_{\ell}(j) \text{ for all } \ell \neq 1
\end{equation*}

Let $\tilde I_{12} = I_{12} \setminus \{j\} \cup \tilde j$ and $I'' = I' \setminus \{k\} \cup \tilde k$. Note that $I''$ is a complement of $\tilde I_{12}$ at $d^*_{11}$.  The following claim holds by construction.

\begin{clm}
$I'$ and $I''$ doesn't have any $(2,1)$ individuals. Moreover, $I''$ has $1$ group $(1,1)$ individual.
\end{clm}

Let $\bar I = I_{12} \cup I' \cup \{\tilde j, \tilde k\}$. First, note that $I_{12} \cup I' \in C(\bar I)$ and $\tilde I_{12} \cup I'' \in C(\bar I)$, since  $M(I_{12} \cup I') = M(\tilde I_{12} \cup I'') = d^*_{11}$. Moreover, let $\hat k$ denote an individual with $\theta(\hat k) = \theta(k)$ and $s(\hat k)< s_K$. Let $\bar I_{\hat k} = \bar I \cup \{\hat k\} \setminus \{k\}$

\begin{lem}\label{case2lemmags}
There doesn't exist an $I^* \in C(\bar I_{\hat k})$ such that $I_{12} \subset I^*$.
\end{lem}
\begin{proof}
Since $\tilde I_{12} \cup I''$ is available and optimal, $I^*$ must also be optimal. For a contradiction, suppose that such an $I^*$ exists. First, note that $\hat k \not \in I^*$, as $\tilde I_{11} \cup I'' \succ J$ for all $J$ that includes $\hat k$ by within-group responsiveness.

Moreover, $\tilde j \not \in I^*$, as otherwise $I^*$ would have $m_{11}$ $(1,1)$ individuals and $m_{21}+1$ $(2,1)$ individuals.
However, this means that $I^* = I_{11} \cup I''$. But then $I^*$ has $m_{11}+1$ $(1,1)$ individuals, which contradicts the optimality of $I^*$.
\end{proof}
As $I_{11}$ is chosen from $\bar I$, but not from $\bar I_{\hat k}$, the result follows.
\end{proof}
Proposition \ref{incompatibilitywithscores:prop} is proved by making the appropriate adjustments to the proof of Proposition \ref{incompatibility:prop}, where decreasing the scores of a set of individuals mirrors the effect of removing those individuals. It shows that whenever the institutions values higher scoring individuals and has a non-trivial preference for diversity, not considering intersectionality when evaluating diversity will cause failure of the gross substitutes condition.

\subsection{Privilege Monotonic Choice Functions}\label{app:privilege}

Propositions \ref{prop:rationality} and \ref{prop:monotonicity} consider the identities of individuals without making affirmative action motives explicit. However, in many cases, the goal of affirmative action policy is tailored to give privileges to certain underrepresented groups. For example, students applying to universities in Brazil can claim certain \textit{privileges} such as ``public HS privilege,'' ``low-income privilege,'' and ``minority privilege,'' where being a member of one of these groups should never hurt a student \citep{aygun2021college}.

To map my model to their setting, I assume that each dimension corresponds to a privilege, and $\Theta_l = \{\theta_l^{\mathcal P},\theta_l^{N}\}$, where $\theta_l^{\mathcal P}$ is the group that can claim the privilege and $\theta_l^{N}$ is the group that cannot claim the privilege.

$I  >_{\mathcal P} I'$ if $\tau(I) \neq \tau(I')$ and there exists a bijection $h: I \to I'$ such that for all $i$, $s(i) = s(h(i))$ and $\theta_l(i) \neq \theta_l(h(i))$ if and only if $\theta_{l}(i) = \theta^{\mathcal P}$. In words, starting with $I'$ and adding privileges to some students, we reach $I$. $\succeq$ is privilege monotonic if $I  >_{\mathcal P} I'$ implies $I \succ I'$. 

\begin{defn}
$I_1,\ldots,I_n$ is a score-choice-privilege (SCP) cycle if 
\begin{itemize}
    \item for each $i<n$, either (i) there exists an $\hat I_i$ such that $I_i \in C(\hat I_i)$ and $I_{i+1} \subset \hat I_i$,  (ii) $I_i >_{\mathcal S} I_{i+1}$ or (iii) $I_i >_{\mathcal P} I_{i+1}$.
    \item either (i) there exists  $\hat I_n$ such that $I_n \in C(\hat I_n)$, $I_{1} \subset \hat I_n$ and $I_{1} \not \in C(\hat I_n)$, or (ii) $I_{n} >_{\mathcal S} I_{1}$ (iii)  $I_{n} >_{\mathcal P} I_{1}$.
\end{itemize}
\end{defn}

SCP cycles characterize the class of choice functions are rationalizable with a utility function that is increasing in scores and privilege monotonic.

\begin{prop}
$C$ doesn't admit a score-choice-privilege cycle if and only if there exists a function $U$ that rationalizes $C$, is wihtin-group responsive and privilege monotonic.
\end{prop}
\begin{proof}
    Follows from replacing $>_{\mathcal S}$ with $>_{\mathcal S} \cup >_{\mathcal P}$ in the proof of Proposition \ref{prop:monotonicity}.
\end{proof}

\subsection{Within-group Responsiveness and Scores}\label{app:extensionscores}

In the main text, I have assumed that $\mathcal S \subset \mathbb R$ and the primitive preference between individuals of the same identity used in the definition of within-group responsiveness is derived from their scores. In this section, I drop the assumption that $\mathcal S \subset \mathbb R$ and use a modified definition of within-group responsiveness that does not depend on the scores and is identical to the definition of \cite{roth1985college} up to the qualifier on groups.

\begin{defn}
     $\succeq$ is \textbf{within-group responsive*} if for all $i$, $i'$ with $\{i,i'\} \cap I = \emptyset$, $\theta(i)=\theta(i')$ and $i \succ i'$, we have $i \cup I \succ i' \cup I$.
\end{defn}

Within-group responsive* does not place any restrictions on $\succeq$ restricted to singleton sets. The preference relation $\succeq$ may even treat $\mathcal S$ differentially based on the identity of individuals. For example, for $s \neq s'$ and $\theta \neq \theta'$, we can have the following:  $i_s^{\theta} \succ i_{s'}^{\theta}$ and  $i_{s'}^{\theta'} \succ i_s^{\theta'}$, where subscripts denote the scores, and superscripts denote the identities. Indeed, the representation in Proposition \ref{prop:rationality} allows for such preferences. The following assumption makes sure the way the institution evaluates scores does not depend on the identities of individuals.

\begin{defn}
    A preference relation $\succeq$ \textbf{treats scores uniformly across groups} if for all $i$, $j$, $i'$ and $j'$ with $\theta(i) = \theta(j)$,  $\theta(i') = \theta(j')$, $s(i) = s(i')$ and $s(j) = s(j')$, $i \succeq j$ implies $i' \succeq j'$.
\end{defn}

When $\succeq$ treats scores uniformly across groups, we can reorder the scores in the following way. Fix a $\theta$ and let for each $s \in \mathcal S$, let $i_s$ denote the individual with score $s$ and identity $\theta$. As $\succeq$ is a preference relation, there exists at least one (but possibly more) $s$ such that $i_s \succeq i_{s'}$ for all $s'$. Let $S_1$ denote the set of such scores. Define $\tilde S$ by assigning all $s \in S_1$ to a real number $\tilde s_1$. We continue in this fashion until the scores are exhausted, where for each $j \geq 2$, define $S_j$ inductively as the set of all scores in $S_{j-1}$ such that $i_s \succeq i_{s'}$ for all $s' \in \mathcal S \setminus \cup_{k < j} S_k$. At each step, we assign all $s \in S_j$ to a real number $s_j$ strictly smaller than $\tilde s_{j-1}$ and create $\tilde{\mathcal S}$. Observe that as  $\succeq$ treats scores uniformly across groups, we would have obtained same order regardless of $\theta$.

Let $\tilde s$ denote the score function that returns the score of each individual on $\tilde S$, $\tilde t$ denote the type function formed by $\tilde s$ and $\theta$, and $\tilde \tau(I)$ denote the function that returns the types of all individuals in $I$ using $\tilde t$ instead of $t$.  $\tilde \succeq$ denotes the preference relation that represents $\succeq$ on $\tilde S$. Formally, $\tau(I) \succeq \tau(I')$ if and only if $\tilde \tau(I) \succeq \tilde \tau(I')$. Note that $\succeq$ and $\tilde \succeq$ are equivalent in the sense that $I \succeq I'$ if and only if $I \tilde \succeq I'$.

\begin{prop}
   If $\succeq$ treats scores uniformly across groups and is within-group responsive*, then $\tilde \succeq$ is within-group responsive on $\tilde{\mathcal S}$.
\end{prop}
\begin{proof}
    Suppose that $\succeq$ treats scores uniformly across groups and is within-group responsive* and take $i$, $i'$ with $\{i,i'\} \cap I = \emptyset$, $\theta(i) = \theta(i')$ and $\tilde s(i) > \tilde s(i')$. As  $\tilde s(i) > \tilde s(i')$ and  $\theta(i) = \theta(i')$, our construction of $\tilde s$ requires $i \succ i'$. As $\succ$ is within-group responsive*, for all $I$ with $\{i,i'\} \cap I = \emptyset$, we have $i \cup I \succ i' \cup I$, which implies $i \cup I \tilde \succ i' \cup I$.
\end{proof}

Note that the converse of the result is immediate as if $\succeq$ is within-group responsive, then by defintion it treats scores uniformly across groups and is within-group responsive*. This shows that although it is sensible to take the scores as the primitive order that determines the preference over singleton sets and define within-group responsiveness accordingly, a more direct adaption of responsiveness to my setting would still work as long as $\succeq$ treats scores uniformly across groups.

\end{appendices}

\end{document}